\documentclass[a4paper, 10pt]{article}

\usepackage{hyperref}
\usepackage{amsfonts, amsmath, amssymb}
\usepackage{graphicx, natbib}
\usepackage{natbib}

\newcounter{takehome}

\title{Exploring the landscapes of ``computing'': digital,
  neuromorphic, unconventional --- and beyond}

\author{Herbert Jaeger}
\date{
  \small{Bernoulli Institute for Mathematics, Computer Science and Artificial
  Intelligence, University of Groningen, and \\
Cognitive Systems and Materials Center (CogniGron), University of Groningen}}

\begin{document}

\maketitle

\begin{abstract}
  \emph{NOTE: An extended and carefully revised version of this manuscript has now been published as ``Toward a generalized
  theory comprising digital, neuromorphic, and unconventional
  computing'' in the new open-access journal \emph{Neuromorphic Computing and Engineering} (\url{https://doi.org/10.1088/2634-4386/abf151}).}

  The acceleration race of digital computing technologies seems to be
  steering toward impasses --- technological, economical and
  environmental --- a condition that has spurred research efforts in
  alternative, ``neuromorphic'' (brain-like) computing
  technologies. Furthermore, since decades the idea of exploiting
  nonlinear physical phenomena ``directly'' for non-digital computing
  has been explored under names like ``unconventional computing'',
  ``natural computing'', ``physical computing'', or ``in-materio
  computing''. This has been taking place in niches which are small
  compared to other sectors of computer science. In this paper I stake
  out the grounds of how a general concept of ``computing'' can be
  developed which comprises digital, neuromorphic, unconventional and
  possible future ``computing'' paradigms. The main
  contribution of this paper is a wide-scope survey of existing formal
  conceptualizations of ``computing''. The survey inspects approaches
  rooted in three different kinds of background mathematics:
  discrete-symbolic formalisms, probabilistic modeling, and
  dynamical-systems oriented views. It turns out that different
  choices of background mathematics lead to decisively different
  understandings of what ``computing'' is. Across all of this
  diversity, a unifying coordinate system for theorizing about
  ``computing'' can be distilled. Within these coordinates I locate
  anchor points for a foundational formal theory of a future
  computing-engineering discipline that includes, but will reach
  beyond, digital and neuromorphic computing.

\end{abstract}

\section{Introduction: why this is a good time to rethink ``computing''}
  
Our modern societies thrive on, and are fundamentally shaped by,
digital computing (DC) technologies. There are a number of reasons why
DC could grow into this majestic role:

\begin{description}
\item[Unversality.] Every information processing task that can be
  specified in a formal (first-order logic) description can
  be solved by a digital computer program. This is so because digital
  computers can emulate Turing machines and Turing machines can
  realize general theorem provers \citep{Jaeger19a}.
\item[Transistors and wires.] In mathematical abstraction, digital
  computing reduces to reading and writing 0's and 1's from and into
  hierarchical data structures. It is fully understood how this
  translates into transistor-and-wire based digital hardware
  architectures, and the corresponding microchip design and
  manufacturing technologies have reached astounding degrees of
  perfection. 
\item[Ease of use.] A hierarchy of mutually cross-compilable
  programming languages --- from hardware-specific assembler coding to
  graphical user interfaces for office software --- allows users on
  all levels of expertise to exploit the potentials of digital
  computers.
\item[Computing---cognition match.] There is a prestabilized harmony
  between rational human reasoning and digital computing. A direct
  line of intellectual inquiry leads from Aristotle's syllogistic
  rules of reasoning through Leibniz, Boole, Frege and the early 20th
  century logicians to Turing who in his groundbraking paper
  \citep{Turing36} still spoke of ``computers'' as humans and of
  the physical states of a computing system as states of mind. Writing
  computer programs is just an exercise in clear ``logical'' thinking.
\item[Unified theory.] There is a unified, standardized body of DC
  theory, comprising automata theory, formal languages, the theory of
  computability and complexity, Boolean and first-order logic. This is
  documented in canonical textbooks and taught to computer science
  students in all universities in the same way, providing a
  conceptual and terminological common ground for a worldwide
  community of DC professionals.   
\end{description}

In view of this intellectual transparency and practical empowerment,
it is understandable that today ``computing'' is largely identified
with digital computing (= ``symbolic'' computing, = Turing computability,
= running ``algorithms''). 

But progress rates of DC technologies are slowing
down and seem to approach serious impasses: 

\begin{description}
\item[Energy footprint.] A widely recited estimate\cite{AndraeEdler15}
  claims that about 10\% of the world's energy budget is due to DC
  technologies, with still exponentially rising rates.

\item[Miniaturization.] Thermal and quantum noise and exploding
  investment costs for microchip fabrication may (or might
  not? see \cite{MurmannHoefflinger20}) prevent further downscaling of
  transistors in commercial microchip production --- the ``End of Moore's
  law'' \citep{Waldrop16}.
      
\item[Toxic waste.] Hardware replacement cycles are ever speeding
  up. Electronic waste ``is now the fastest-growing waste stream in
  the world'' \citep{WorldEconomicForum19}.
        
\item[Software complexity.] Software products are ever growing in size
  and complexity, perpetuating the \emph{software crisis} since it was
  first acknowledged in the mid-1960s \citep{Ebert18}. Given that
  critical segments of our modern world become permeated by complex
  software systems, we may be in for ruptures of societal
  functionality.
\end{description}

Such boundary conditions have led to a surge of explorations in
``brain-like'', \emph{neuromorphic computing} (NC)
technologies. ``Learning from the brain'' seems a promising route
toward escaping from some of the DC impasses:

\begin{description}
  \item[Energy efficiency.] Brains need only a minute fraction of the
energy consumed by digital supercomputers for ``cognitive''
tasks \citep{Boahen17}.
\item[Unclocked parallelism.] The inherent, complete parallelism of
  the brain's \emph{in-memory computing} \citep{IelminiWong18} stands
  in stark contrast to the serial processing in DC systems where only
  a fraction of all available transistors are active at any time
  \citep{Pelaez90}.
\item[Cognitive-style computing.] Artificial neural networks (ANNs)
  operate in ways that seem akin to human cognitive
  processing. Besides the fundamental fact that ANNs are not
  programmed but \emph{trained}, they can, for instance, generate
  striking visual art \citep{Olahetal17} or win against human
  world champions in the most cognitively demanding games
  \citep{Silveretal16, Berneretal19}.

\item[Robustness and adaptability.] Biological brains cope well
  individual neuron death and even extensive lesions. Their cognitive
  processing adapts to changing contexts, from reliable object recognition
  in fast-changing lighting conditions to lifelong learning. While machine learning has only begun to
  understand such physical and functional
  robustness \citep{Saundersetal19, Heetal19a, Zhangetal19}, brains are
  living proof that the hardware and functional brittleness of DC
  systems can be overcome.
\end{description}

Investigating neural networks has a long tradition and the field has
gone through hypes in the past --- the most famous one unleashed by
the invention of the Perceptron \citep{NYT58}, followed by a less
spectacular one around 1990 when the backpropagation algorithm made
the training of (not too deep) multilayer feedforward networks broadly
applicable.  There are however indications that the current flush of
interest in NC has sustainable foundations:

\begin{description}
\item [The deep learning revolution] \citep{TuringAward18}
  has manifested the powers of artificial neural information
  processing to scientists, decision-makers and the general public
  alike.
\item [Large-scale digital neuromorphic microchips,] developed by the
  leading microchip manufacturing companies, emulate neural spiking
  for low-energy implementations of neural networks
  \citep{Merollaetal14, Daviesetal18, Neckaretal19}. This appears
  as a visible proof of the economical potential of MC.
\item[Neuro-inspired algorithms] \citep{LukoseviciusJaeger09,
    FremauxGerstner16} have been deployed on (partially) non-digital
  hardware \citep{IndiveriLiu15, Yousefzadehetal18, Tanakaetal18,
    Neckaretal19}.
\item [Artificial neural retina and cochlear implants] partially
  restore vision and hearing \citep{Chuangetal14, Lenarz17}.
\item [Spiking neural camera sensor chips] with integrated neural
  processing yield ultrafast computer vision
  \citep{Gallegoetal20}.
\item [A neuro-optical internet communication link] prototype is entirely passive and 
  needs no external energy supply \citep{Freibergeretal17}.
\item [The advent of memristors] in NC microchips
  \citep{YangStrukovStewart13} has spurred material scientists to
  explore a wide range of physical nanoscale phenomena for
  computational exploits \citep{Coulombeetal17, Torrejonetal17,
    Prychynenkoetal18, Chenetal20, Miriglianoetal20, Prucnaletal20}.
\end{description}

However, the neurosciences do not yet provide readily implementable
blueprints for engineering computing systems. How the brain
``computes'' is understood only in fragments. Foundational questions
remain debated. For example, it is not settled how ``symbols'' or
``concepts'' (addressable, stable representational entities) emerge
from neural dynamics \citep{Gross02, Durstewitzetal00, Baddeley03,
  LinsSchoener14, Jaeger17, Besoldetal17, Werneckeetal18}; how several
such entities are coupled into composite entities
\citep{BuzsakiChrobak95, SlotineLohmiller01, Legensteinetal16};
how information streams can be dynamically routed in a brain
\citep{Olshausenetal93, Hoerzeretal14, Sabouretal17}; or how and
in what sense ``information'' is encoded in neural dynamics
\citep{Gerstneretal97, Panzerietal17}.

Stepping back from the daunting complexity of concrete biological
 brains, one may ask a meta question which only at first sight
looks naive: how could Nature ever ``invent'' such magnificient
systems? --- For eons, biological evolution has been discovering,
differentiating, optimizing and cross-coupling myriads of different
biochemical, electrophysiological and anatomical phenomena,
integrating them into that supremely adaptive, robust and balanced
physical system that we carry in our heads. The structural, dynamical
and functional complexity of this system's organization spans 
many orders of magnitude of spatial and temporal scales. Yet,
throughout this breathtaking complexity, there is one grand unifying
boundary condition: whatever phenomenon is exploited in a brain, it
arises from the biochemistry and electrophysics of wet, soft
biological tissue. Biological brains \emph{must} use only that which
is physically possible in a biological substrate --- and they
positively \emph{do} use that.

Current artificial NC microsystems are, and future ones likely will
be, manufactured from more enduring materials. Furthermore,
engineers are already active to exploit physical effects that cannot
occur in biological tissue, for instance optical, ferroic, skyrmionic,
or even micromechanical effects \citep{Prucnaletal20,
  Everhardtetal19, LeonovMostovoy15, Coulombeetal17}. 

If (i) a grand lesson to learn from Nature's brains is to exploit just
everything which the available physical substrate offers, and if (ii)
future hardware substrates will differ substantially from biological
tissue, then it makes all sense to considerably widen the neuromorphic
computing agenda, exploring how whatever physical phenomena in
whatever material can be harnessed for ``computing''.

Exploiting ``the physics of the materials directly''
\citep{Zauner05} is absolutely not a new idea \citep{Zauner05,
  EuropeanCommission09, StepneyHickinbotham15a, Horsman17,
  Adamatzky17a, Adamatzky17b, StepneyRasmussenAmos18}. This theme has
been investigated for decades from different angles under a diversity
of namings --- for instance \emph{unconventional}, \emph{natural},
\emph{emergent}, \emph{physical}, \emph{in-materio} computing, ---
sometimes evoking a strong echo like DNA computing
\citep{vanNoortetal02}, sometimes rather restricted to an academic
niche like computing with fungi \citep{Adamatzky18}.  A variety of
classification schemes have been proposed \citep{Harnad94,
  deCastro06, BurginDodigCrnkovic13, Stepney17} for approaches in
the unconventional computing (UC) research landscape. However, a unifying
theoretical framework does not yet
exist. \cite{StepneyHickinbotham15a} list a number of existing
mathematical formalisms that are tailored to specific subsets of
material phenomena or computational functionalities, and otherwise
remark that an \emph{``over-reaching formalism ... may be
  desirable''}. Even when the understanding of ``computing'' is
confined to the digital-symbolic paradigm, a general theory of
computing in non-digital substrates is desired but missing
\citep{Horsman17}.

To avoid misunderstandings I point out three things that I do
\emph{not} think of when I speak of exploiting physics
directly. First, I am not concerned with \emph{pancomputationalism}
where formal concepts from symbolic computing are invoked to describe
and explain the physical world \citep{Lloyd13}. Second, I am not
dealing with \emph{physics of computation}, a field that explores the
physical boundary conditions of digital computing
\citep{Wolpert15}. While these researches give inspirations for
mathematical formalizations of complex, self-organizing, physical
pattern formation \citep{Zuse82, Wolfram02, Fredkin13}, one must
be aware that these traditions are immersed in the DC understanding of
``computing'' as discrete symbol manipulation processes. Finally, I
see the wider fields of unconventional computing as hardly
intersecting with \emph{quantum computing}, which has already matured
into a discipline of its own standing.

Let me summarize all these observations, adding my personal opinion:

\begin{enumerate}
\item The DC paradigm and the technologies arising from it define the
  standards, formal models, intuitions and expectations that shape our
  concept of ``computing''. Only recently, flattening rates of
  progress and growing environmental and societal concerns have
  prepared the grounds for substantial investments into alternative
  paradigms of ``computing''.
\item Neuromorphic computing is currently the most energetically
  investigated alternative route to ``computing''.  However, despite
  manifold promising initial achievements, swift and broad progress is
  hampered by a fragmentation of the field and the absence of a
  unifying theoretical foundation.
\item While NC is guided by the ``learning from the brain'' rationale,
  there is a long history of propositions to establish uncoventional
  computing paradigms which shortcut the brain role model and aim at
  exploiting whatever physics can offer directly. Like in the case of
  NC, a formal theoretical foundation for UC is missing.
\end{enumerate}

In this situation I venture a rather daring hypothesis:

\fbox{\parbox{13.6cm}{
  It is possible to generalize the theory of symbolic
  computing, by
  \begin{itemize}
  \item generalizing from the physical \emph{switchable bi-stability} of
    digital transistors to a much wider class of \emph{modulatable
      dynamical modes} of novel nonlinear devices, and
  \item generalizing from the 0 - 1 (or true - false)
    \emph{symbolic} abstraction of physical bi-stability to a suitable
    \emph{qualitative} abstraction of modal nonlinear dynamics,
  \end{itemize}
  leading to the development of a unified and comprehensive theory for
  \emph{modal computing} (MC) which
  \begin{itemize}
  \item enables the principled exploitation of novel, non-digital
    nonlinear materials and devices for ``computing'',
  \item contains DC as a special case,
  \item offers a basic perspective to analyse biological neural
    dynamics and design MC hardware, circuits and architectures, and
  \item unifies existing UC approaches.
  \end{itemize}}}

As one may suspect, I cannot presently offer a worked-out definition
of dynamical modes or their qualitative abstraction. Nonlinear
dynamical systems can exhibit an unlimited richness of ``behavioral''
\citep{AbrahamShaw92} phenomena, and the question \emph{which} of
them should be focused and \emph{how} they should be qualitatively
characterized is, at this time, a question unanswered. 
In another paper (in preparation) I attempt to develop
intuitions and starting ideas for approaching this question.

Here I want to investigate a broader, non-mathematical, real-world
question which also needs to be well understood if one wishes to
establish a generalized engineering science of ``computing''.

The powers of the DC paradigm do not emerge from a single, closed formal
theory. Besides the model and theory of Turing machines which could be
(mis)taken as ``the'' fundamental theory of DC, there are other
formalisms, models and theories that are just as essential for the
real-life manifestations of DC. They include the theories of automata,
formal grammars and languages, programming languages and compiler
design, computability and complexity theory, Boolean and first-order
logic, and metalogical frameworks. Only the totality of
these formalisms, models and theories instruments
the real-world concerto of professional DC activities, from
device engineering to microchip fabrication technologies, from circuit
design to computer architectures and communication networks, from
programming language development to human-computer interfacing, from
databases to internet services,  from
beginners' programming exercises to software engineering and use-case
specification frameworks, and all the rest. 

One of the challenges faced by NC is that such a comprehensive theories
(plural) ecosystem needs a longer time to grow than this young field
so far has had. As a consequence, large and always renewed ad-hoc
efforts are still needed to transfer any single novel NC technique
from lab A to lab B, let alone to a wider user community. Our own
experiences in this regard shine through every line of our project
report in \cite{Heetal19}. High-investment efforts to define systematic
multi-level workflows in the NC domain \citep{Zhangetal20}
demonstrate that there is an urgent demand to meet.

The efforts of \cite{Zhangetal20} and theoretical work in the UC
domain \citep{Horsman17} are still embedded in the original DC
conception of ``computing'' as executing ``algorithms'', which upon a
finite input run a finite time --- during which they are decoupled
from input --- and return a finite output. I believe that a
generalization from digital to modal computing will have to include
online processing scenarios. A prime reason for this belief is that
this is the principal mode of operation for biological brains. A
second reason lies in the nature of dynamical modes, which likely will
turn out to be often transient and entrained to a stream of input
signals. Another important difference between DC and MC is that
digital computers can be \emph{programmed}, whereas it may turn out that most
MC systems need to be \emph{trained}. Yet another difference is that
DC microchips come in functionally identical copies, while MC hardware
systems will not always be identically reproducible due to device
mismatch. They might require individual training, leading to
individual use life histories. All of this will make an MC theory
ecosystem look and function profoundly different from what we know
from DC.

\fbox{\parbox{13.6cm}{
  The question that I address in the remainder of this article: what
  sorts of sub-theories and models are needed for any full-fledged
  engineering discipline so that it can claim to be a ``computing''
  discipline?}}

\section{Staking out the ``computing'' landscape }

In order to reveal options for possible answers to this question, I
will try to work out universals and essentials across several
existing ``computing'' paradigms besides DC, including views from NC
and UC.

This is a very long section. At the end I provide a summary (Section
\ref{secSummary}) which should be more or less self-contained, such
that readers can directly jump forward to it.

\subsection{Approach}

\begin{figure}[htb] 
\centering
\includegraphics[width=15.5cm]{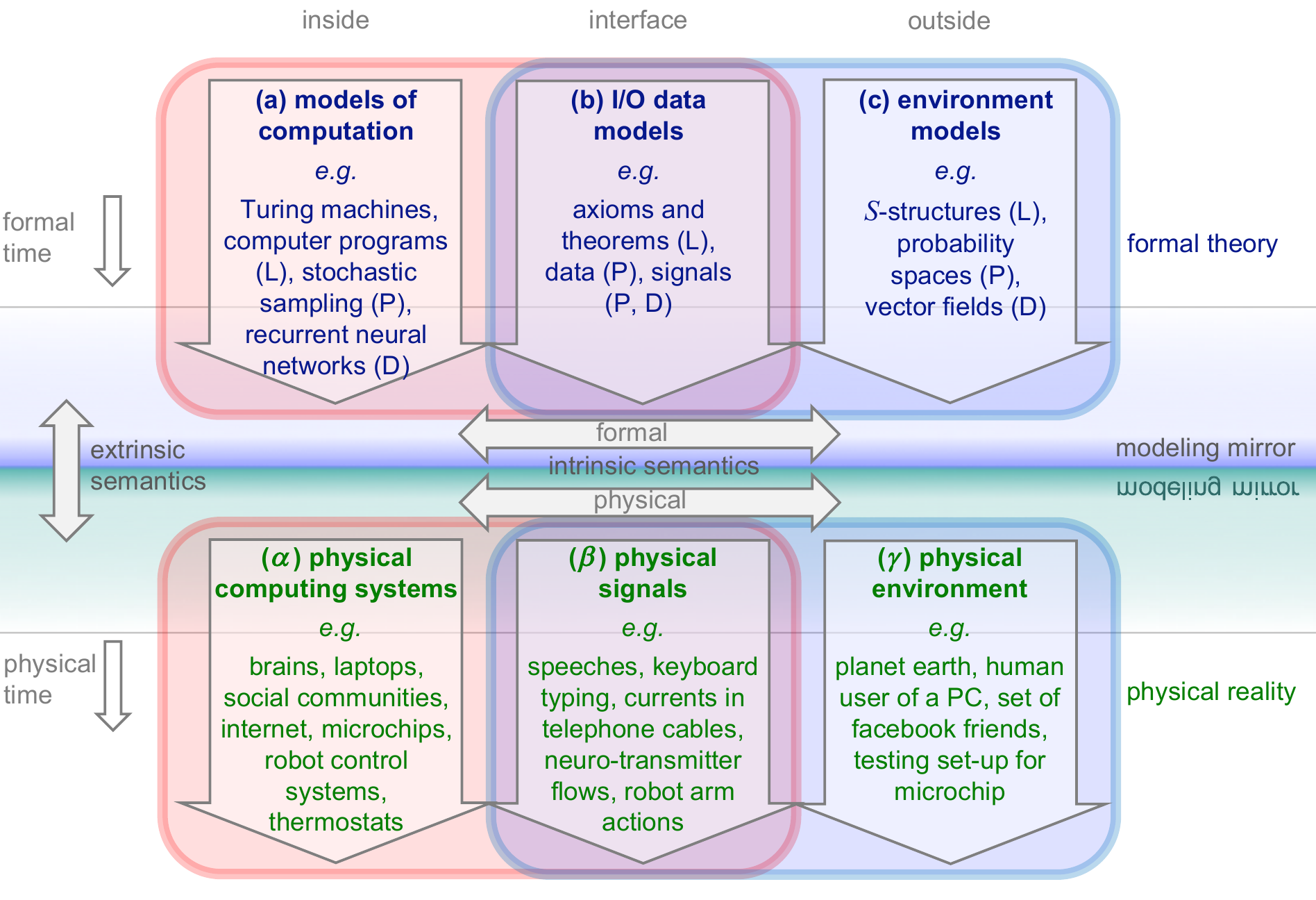}
\caption{Schema of the six main elements \textbf{a}, \textbf{b},
  \textbf{c} [formal] and \textbf{${\alpha}$}, \textbf{$\beta$},
  \textbf{$\gamma$} [physical] in my navigation map
  to locate formal theories, which here are loosely characterized as
  logic-based (L), probabilistic (P) or dynamical systems oriented
  (D).}
 \label{figGrandSchema}
\end{figure}

Figure \ref{figGrandSchema} shows the main conceptual components that
I want to explore in some detail and to relate to each other. I want
to dissect the physical reality (bottom half of figure) into the
computing hardware systems \textbf{(${\alpha}$)} and the physical
environment \textbf{(${\gamma}$)} they are embedded in and in which
they should serve some purpose. The interface boundary
\textbf{(${\beta}$)} between these two comprises the physical signals
that are exchanged between the computing systems and their
environment. These three segments of physical reality are mirrored in
the non-physical domain of mathematical abstraction by corresponding
formal representations of the physical computing system, providing
information processing models \textbf{(a)}, input / output (I/O) data
models \textbf{(b)}, and formal models of outward physical realities
\textbf{(c)}. Figure \ref{figGrandSchema} fills each of these six
component boxes with a number of suggestive examples.

Given a specific computing system with its environment and/or the
formal models thereof, it is a matter of convention where the boundary
interface is put. For instance, when discussing human brains,
the visual sensory input signal boundary could be placed at the
interaction between photons and photoreactive receptor molecules
inside retinal photocells, or it could be assigned to the spike
trains sent through the optical nerve.

One of the questions that must be implicitly or explicitly answered by
any MC theory is what makes a MC system ``compute''. The classical
answer, which is also employed in all accounts of unconventional
computing that I am aware of, is to define ``computing'' in the
footsteps of DC. This conceptualization involves several steps, namely
first encoding computational ``tasks'' into a formal input, which is
then physically entered into a physical computing system (a step that
passes the \emph{modeling mirror} between abstract formalism and
physical reality and involves un-formalizable pragmatics), then let
the physical system physically evolve until some physically observable
halting condition is met, then read off the physical system's state
into a formal output (again, this step crosses the modeling mirror and
involves pragmatics), then decode the output into the task's
solution. I do not think that this view of ``computing'' is suitable
for MC, one reason being that biological brains (which an MC theory
should cover as special cases) operate in a continuous temporal
input-to-action interaction with the environment which is quite
different from the time-decoupled single-input-to-single-output
transformation functionality expressed in the classical view. I
believe that brain-like (or more generally, many modal computing)
systems will often be operating in a modality where they are
\emph{entrained} to a driving input stream. This is different from DC
accounts of online processing, where the processing system is
executing a series of ``tasks'' fast enough to stay on track with the
incoming processing demands, like in the model of interactive Turing
machines \citep{vanLeeuwenWiedermann01}. During the computational
operations needed to solve the individual tasks however the processing
machine decouples itself from incoming input. A more general view of
``computing'' is needed which natively covers entrained input stream
processing.

While I cannot give a complete set of necessary and
sufficient conditions, I think that the following four are necessary
and quite informative for a physical system to be called ``computing''
in an defendable way:
\begin{enumerate}
\item ``computing'' requires a \emph{temporal evolution} of the system
  that computes --- this is why the temporal arrows 
  are so prominent in the figure;
\item ``computing'' involves that ``information'' feeds into, and comes
  out of the computing system: computing systems must be \emph{open}
  systems --- this is why I devote a special conceptual place
  \textbf{($\beta$)} to the interface boundary between the computing
  system and its environment;
\item ``computing'' operations going on in a computing system should
  be \emph{cognitively interpretable}, being  relatable to
  some aspect of human cognition --- this is the heritage of the
  history from Aristotle to Turing (not pictured in Figure
  \ref{figGrandSchema});
\item ``computing'' is \emph{meaningful}, that is, some
  \emph{semantic} account of ``what'' is computed should be possible
  --- indicated in the figure by the two semantics arrows (to be
  detailed below).
\end{enumerate}

A formal theory needs a mathematical language to be expressed. What
mathematical language is used is often tied to the primary
disciplinary background of the modeler. Researchers with a background
in computer science or AI tend toward discrete-algebraic and
logic-based formalisms; cognitive scientists may prefer probabilistic
(Bayesian) frameworks; physicists tend toward dynamical systems based
modeling languages. These are marked L, P, D (for Logic, Probability,
Dynamical systems) in Figure \ref{figGrandSchema}. The adopted
background mathematics determines which phenomena can be modeled and
which not. The chosen background mathematics often reflects a
deep-rooted personal or community perception of physical or cognitive
reality. This may evoke hefty epistemological debates, like the
decades-long \emph{physical symbol systems hypothesis} controversy
about the physical-neural reality of ``symbols'' in cognizing brains
\citep{NewellSimon76, FodorPylyshin88, Brooks91b, PfeiferScheier98,
  LairdLebiereRosenbloom17, Buongiorno19}. Entire modeling schools
(like cybernetics, classical AI, or Bayesian agent modeling) have
grown around a fundamental commitment to a specific background
mathematics, and have helped to advance the very mathematical methods
in return. In the final section I will argue that neither L, P or D
type mathematics alone is appropriate for MC.

The two directions bottom-top and left-right in Figure
\ref{figGrandSchema} can be regarded as semantic axes. The bottom-top
axis, which I will call \emph{extrinsic} semantics, connects formal
models with their physical counterparts, crossing the modeling mirror
between the ontological domains of the formal and the physical. Since
the physical end of this axis is just the physical reality, extrinsic
semantic pairings between models and reality cannot be
formalized. Instead, this link is established by social conventions
and practical routines. Students learn to connect the two ontological
domains by practical exercises and epistemologists work out, using
plain English, what it means that a formal model is ``valid'' or
``adequate'' or even ``true'' in some sense or other. The left-right
axis is the bridge between what happens or is ``represented'' inside a
computing system and what happens or ``exists'' in its outside
environment. This bridge is instantiated twice, within the formal
domain where it connects segments \textbf{a} and \textbf{c}, and
within the physical domain where it connects $\alpha$ with
$\gamma$. Because these semantic relations are spanned within their
respective ontological domains I will call them \emph{intrinsic}
semantics. Formal intrinsic semantics can be mathematically formalized
because on both ends \textbf{a} and \textbf{c} of the formal intrinsic
semantic arrow there are mathematically defined constructs.  This will
be an important theme in the detailed discussion given below. The
physical intrinsic semantics comprises all the real-world interactions
between a physical computing system $\alpha$ and its physical
environment $\gamma$, mediated by physical signals. For example, what
happens inside a thermostat (in segment $\alpha$) interacts in
physically lawful ways with what happens in the room (in segment
$\gamma$) where it has been installed.

The study of semantic relations between symbols and their ``meaning''
is an ancient and richly thought-out theme in philosophy. I am only
superficially acquainted with this body of thought and my separation
of semantic relationships into extrinsic and intrinsic ones is
certainly a stark simplification, and my division of everything into
merely two ontological domains (formal and physical) is naive. It
seems to me however that in much of the philosophical thinking, the
role of of formal models of physical realities is ignored. I found the
simple ontological bipartition of everything and the resulting
intrinsic / extrinsic separation simply quite helpful in organizing my
thinking, and I make use of it when I explain
logic to computer science students \citep{Jaeger19b}.

The schema shown in Figure \ref{figGrandSchema} suggests a simplicity
than does not exist. Historical theory-building processes are in many
respects like biological evolution, leading to different solutions in
different niches (scientific disciplines and
communities). Methodological differences can be drastic, as between
the humanities and the natural sciences. But even within and between
the neighboring disciplinary strands of computer science, AI, and
machine learning we find distinctively different ways of formalizing
and theorising about ``computing''. Before one embarks on MC theory
development it is instructive to inspect existing ``computing''
theories and formalisms at a higher resolution than in Figure
\ref{figGrandSchema}.

The words ``theory'', ``formalism'' and ``model'', which I will use a
lot, should be handled with care. I understand them in the following
way. A \emph{formalism} is a set of more or less rigorously specified
conventions of what kind of formal expressions one may write
down. Examples are the formalism of first-order logic, the formalism
of probability theory, or the formalism of ordinary differential
equations. Also every programming language is a formalism. Formalisms
are languages. Logicians and computer scientists often specify
formalisms in complete accuracy through grammars, while mathematicians
in general and physicists rely on informal conventions (which
nonetheless are binding and well-understood). A \emph{model} is a
specification of a particular piece of reality, written down with the
notational tools of a specific formalism. Finally, in the strict sense
of mathematical logic, the word \emph{theory} denotes the set of all
theorems that can be proven from a set of axioms, like the theory of
groups. Outside logic, the word is used in a wider and less strict
sense, often comprising an entire collection of intended models,
specifications of observation procedures, and predictions and
hypotheses, like in the theory of quantum mechanics or the theory of
supply and demand. I generally adopt this second usage and say
``deductive theory'' when I mean the first one.

A complete coverage of all formalisms and models in the symbolic-logic
(L), probabilistic (P), and dynamical systems oriented (D) modeling
domains is infeasible and I have to confine the discussion to selected
fragments of the total modeling cosmos. The theory landscape in
symbolic/digital computing is canonically worked out and its
essentials fit into one textbook which all students of computer science
indeed have to digest. I will be able to give an almost
comprehensive account here. This is not the case for the P and D
domains whose theory-building landscapes are very diverse and
unifying meta-views are not available. In the
probabilistic modeling domain I will consider only formalisms and
models where probability distributions are represented and
``computed'' through sampling mechanisms, and use the acronym SPPD
(``sampling-based processing of probability distributions'') to
characterize this kind of formalisms and models. With regards to dynamical
systems modeling I will only inspect the fragment where ordinary
differential equations (ODEs) are used and speak of ODE formalisms and
models.

Within each of the L, P and D domains I will find it helpful to
distinguish between two kinds of formalisms and models.

The first kind
captures what one could call the ``mechanics'' of computing
processes. I will call these \emph{how-}formalisms and models. In the
three domains this will be the abstract models of algorithms and the
programming languages (L), sampling algorithms and physics-oriented
models of certain stochastic processes that are interpreted as
sampling processes (P), and ODE models of brains and analog computing
machines (D).

\begin{figure}[htb]  
\centering
\includegraphics[width=14cm]{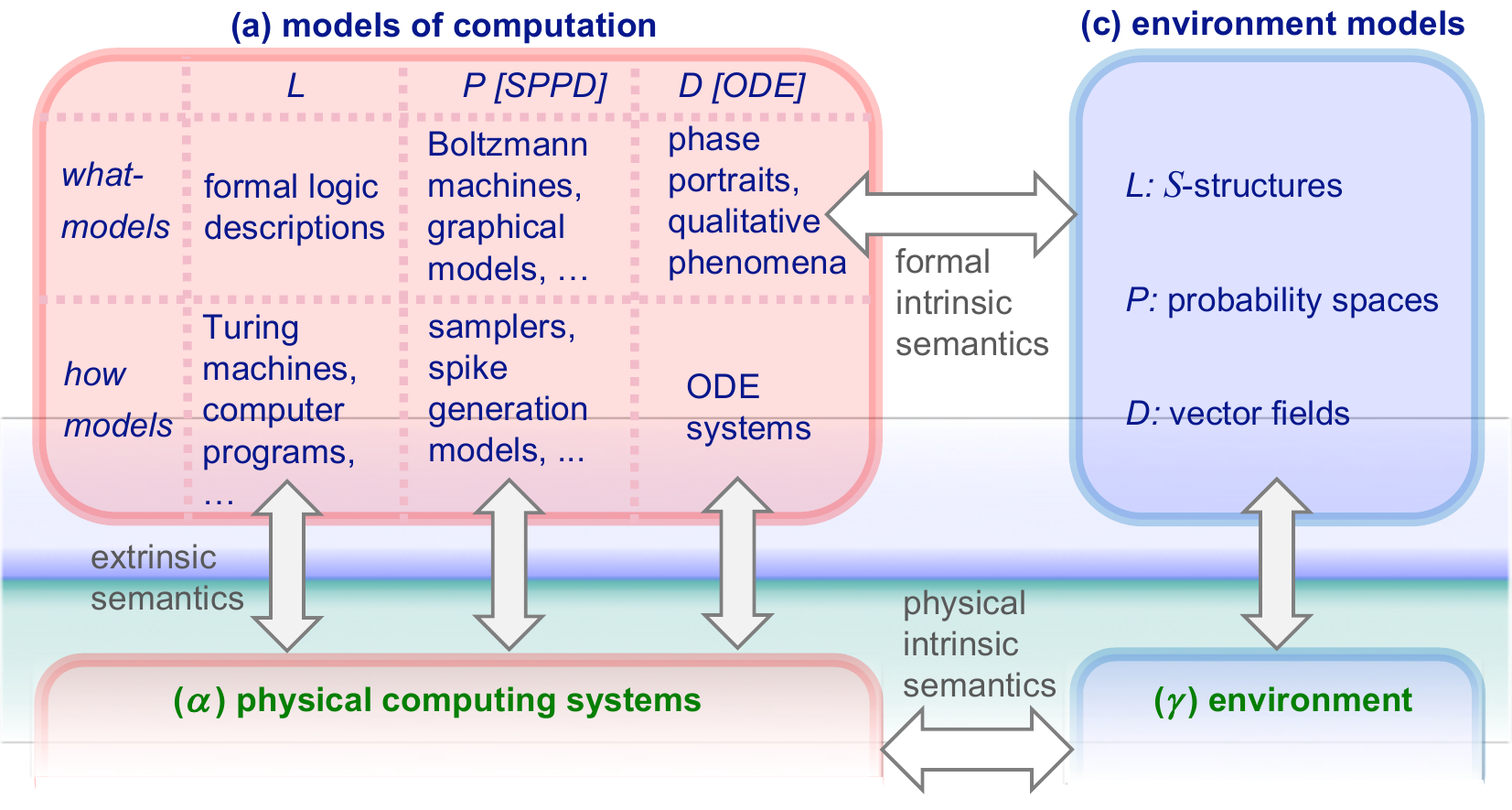}
\caption{Zooming into Figure \ref{figGrandSchema}, giving an overview
  of themes treated in my survey of theories
  of computing conceived in the perspectives of symbolic/logic (L),
  sample-based probabilistic (P [SPPD]), and ODE-based dynamical
  systems (D [ODE]) modeling.}
\label{figZoomIn}
\end{figure}

The second kind comprises formalisms and models whose core constructs
capture cognitively interpretable aspects of human information
processing. I will refer to these as \emph{what-}formalisms and
models. In the L domain this will be logical inferences on the basis
of symbolic configurations, and the what-formalisms are the formalisms
of symbolic logic, of which there are many. In the P domain the core
cognitively interpretable constructs are probability distributions. In
probabilistic accounts of cognitive processing (in theoretical
neuroscience, cognitive science, AI and machine learning), probability
distributions are widely considered as a useful and appropriate
mathematical correlate of \emph{concepts}.  Probabilistic what-models
there are as many as there are ways to conceive probabilistic
conceptual reasoning, but all of these models are expressed in a
single formalism, namely the canonical textbook formalism of
mathematical probability theory. Finally, in the D domain, the
cognitively interpretable mathematical constructs that can be defined
on the basis of ODEs are qualitative geometrical constructs like
\emph{fixed points}, \emph{attractors}, \emph{bifurcations} and (in
nonstationary dynamics) dynamical \emph{modes}, and many more.  In the
cognitive and neurosciences, such geometrical items have become
perceived as correspondents of a diverse range
of cognitive phenomena. Similar to what we find in probabilistic
modeling, what-models there are many, but they are all written down
using the same basic ODE formalism. Figure \ref{figZoomIn}
shows a zoom into segment \textbf{a} of Figure \ref{figGrandSchema}.

I will discuss L, P, D formalisms with regards to the themes
\begin{enumerate}
\item interrelationships between formalisms,
\item formal intrinsic semantics,
\item formal time, and
\item hierarchical organization of formal constructs within a
  formalism.
\end{enumerate}

I leave a discussion of other, likewise relevant themes for another
occasion. In particular I will not consider aspects of
\begin{itemize}
\item functionality --- what tasks or problems are solved by
  ``computing'',
\item pragmatics --- whether one can ``program'' a computing system or
  one has to ``train'' it; how can ``users'' interact with the
  ``computer'', and what are the ``users'' in the first place,
\item information --- how ``encoding'', ``uncertainty'', ``precision''
  are formalized,
\item learning, adaption, calibration, stabilization,
\item growth --- extensibility of models and physical systems,
\item complexity --- what are measures and limits of a system's
  ``computing power'',
\item intra-model communication --- what are ``signals'' and how are
  they propagated inside a computing systems,
\item formal space --- whether or how are how- or what-models
  metrically or topologically structured, and how this relates to
  formal time.
\end{itemize}

The four themes which I do discuss are key coordinates for theorizing
in digital computer science. But upon inspection of non-digital
information processing models in the P and D domains it will turn out
that these themes are also instructive entry points to study
non-digital computing concepts, and that they can be worked out in
very different ways from what we are accustomed to in DC.

\subsection{Findings}

This subsection constitutes the main substance of this paper. It is 
long. An almost self-contained summary is provided as a separate
subsection at the end.

\subsubsection{DC formalisms and models}

I begin my detailed investigation with a discussion of digital
computing formalisms and models. A first overview: The how-formalisms
include abstract descriptors of algorithmic processing mainly used for
theoretical analysis (like the Turing machine or certain grammar
formalisms) as well as formalisms that are destined for practical use,
namely programming languages. Some abstract how-formalisms are useful
also for practical programming, in particular lambda calculus (which
directly spins off the so-called functional programming languages) and
the random-access machine model (which is close to assembler
programming languages). --- The what-formalisms are logic formalisms,
with Boolean logic and first-order logic being the standard textbook
representatives. What-models are sets of logical formulas that specify
what a program should compute. Such specifications are needed when one
wants to formally verify that a written computer program actually
fulfils its purpose --- an important objective in many application
domains. A special case are the so-called declarative programming
languages, like Prolog or functional programming languages, which can
be dually regarded as how- and what-formalisms and allow the user to
write programs directly in terms of ``what'' tasks shall be computed
by the program.  All of the above are concisely explained in my online
lecture notes on theoretical computer science \citep{Jaeger19b,
  Jaeger19a}.

There are further formalisms that belong to the wider circles of DC
theory which do not fall into the how- and what- classes that I
mentioned here. For example, I exclude from consideration formal
communication protocols between different computers. The formalisms
that I mentioned however can be considered the bone and marrow of
theoretical computer science.

\emph{Take-home message \stepcounter{takehome}\arabic{takehome}: The
  separation between how- and what-formalisms and models is not
  entirely clear-cut. Some how-formalisms can climb to levels of
  cognitively interpretable abstraction that they could also be
  considered what-formalisms.}

\textbf{Formalism interrelations.} Both how- and what-models in DC are
noted down using tools from discrete mathematics, being based on
finite alphabets of symbols. How-models describe computing processes
in terms of step-wise, rule-based updates of compound symbol
configurations (example: the zipper algorithm from Box 3). There is a
well understood hierarchy of how-formalisms, defined by different
levels of \emph{expressiveness}. A formalism $A$ is at least as
expressive as a formalism $B$ if every input-output task that can be
solved with models formalized in $B$ can also be solved with a model
formalized in $A$. This can be equivalently stated in a ``syntactic''
way, as follows. Consider a model $\mathcal{B}$ formalized in $B$
(think of a computer program $\mathcal{B}$ written in a programming
language $B$).  Then there exists a model $\mathcal{A}$ formalized in
$A$ and an encoding function $\tau_{\mathcal{B} < \mathcal{A}}$ which
translates any symbolic configuration $b$ that can occur in a ``run''
of $\mathcal{B}$ into an $\mathcal{A}$-configuration $a$ (that is
$\tau_{\mathcal{B} < \mathcal{A}}(b) = a$) such that, when $b \to b'$
is a configuration update step that can occur when running
$\mathcal{B}$, there is a sequence of update steps in $\mathcal{A}$ of
the form
$\tau_{\mathcal{B} < \mathcal{A}}(b) \to a_1 \to \ldots \to a_n \to
\tau_{\mathcal{B} < \mathcal{A}} (b')$. In technical terminology:
every run of $\mathcal{B}$ can be \emph{simulated} by 
$\mathcal{A}$. This leads to a hierarchy of classes of formalisms,
where within each class, all formalisms have the same expressivity
(they can mutually simulate each other), and where formalisms in less
expressive classes can be simulated by formalisms in more expressive
classes but not vice versa. This hierarchy, often referred to as the
\emph{Chomsky hierarchy}, is standardly explained in computer science
textbooks and can be regarded as the backbone of theoretical computer
science. The most expressive class known today is the class of
Turing-equivalent formalisms. It comprises all practically used
programming languages and many abstract mathematical formalisms,
including of course the formalism of Turing machines. According to the
\emph{Church-Turing hypothesis}, no more expressive kind of how-formalism
exists.

While all Turing-equivalent how-formalisms are equally expressive, it
is nonetheless useful to arrange them, within their class, in an
abstraction hierarchy, with ``low-level'' formalisms close to the
fine-grained 0-1 switching of transistors in digital hardware (for
instance assembler programming languages), and with ``high-level''
formalisms (like declarative programming languages or Microsoft Excel)
closer to the human programmer's cognitive representations of what it
means to process information. When one writes a program in a
high-level language and starts it from a high-level user interface,
behind the scene it becomes level-wise \emph{compiled} into lower-level
ones until a version specified by the microchip manufacturer is
reached, which can directly address the hardware $\alpha$, crossing
the modeling mirror shown in Figure \ref{figGrandSchema}.

Trained computer scientists can easily create new, again equivalent
how-formalisms tailored to their particular needs. This flexibility
and transparency to move up and down in the abstraction hierarchy is
unique in digital computing and is one reason for the current
intellectual and practical dominance of DC over all other approaches
to ``computing''.

\emph{Take-home message \stepcounter{takehome}\arabic{takehome}:
  How-formalisms can be systematically ordered by their
  expressiveness.  A formalism A is more expressive than a formalism B
  when all B models can be simulated by A models. Within the class of
  most expressive formalisms --- the Turing-equivalent ones --- one
  can furthermore roughly order them according to how close they are
  to ``low'' levels bordering to the physical 0-1-switching of digital
  circuits, or to ``high'' levels that are more interpretable in terms
  of human cognitive operations.}

As to what-formalisms (formal logics), we find a wealth of different
logics (plural) in the DC domain. Besides the two logics which are
standardly taught to students of computer science (Boolean and
first-order logic), there is also second-order logic, trimmed-down
fragments of first-order logic, and indeed infinitely many more. A
core theme of contemporary theoretical computer science is to work out
\emph{logical frameworks} \citep{Rabe08} to navigate in this richly
populated landscape. A key ordering principle is \emph{logic
  translations}, that is meta-formalisms which allow one to specify
how one logic A can ``express'' everything that another logic B
can. This is only superficially similar to the expressiveness
hierarchy we saw for Turing-equivalent how-formalisms. The
expressiveness hierarchy for how-formalisms is defined on the basis of
the \emph{syntax} of symbolic configurations, while for logic
translations a \emph{semantical} account is needed of what the symbolic
expressions that can be written down in a logic formalims
mean. This renders the study of expressiveness relations
between what-formalisms much more involved than it is for
how-formalisms. Research in this area is far from being 
canonically completed.

\emph{Take-home message \stepcounter{takehome}\arabic{takehome}:
  What-formalisms in DC are formal logics, of which there are
  many. Like how-formalisms they can be ordered according to their
  expressiveness, but here ``expressiveness'' is defined in terms of
  formal intrinsic semantics, not in terms of syntactically describable
  transformations of symbol configurations.}

\textbf{Formal time} surfaces in distinctively different ways in how-
versus what-formalisms. For how-formalisms the story is quickly
told. The formal, discrete update rules acting on symbolic
configurations become ultimately mirrored in the physical clock cycles
of digital microchips. For what-formalisms it is more difficult to
understand time. Explaining how time arises in what-formalisms is
intimately connected to \textbf{intrinsic semantics}. What-formalisms
cast and connect the segments \textbf{a}, \textbf{b}, \textbf{c} from
Figure \ref{figGrandSchema} with the mathematical constructs of formal
logic, often first-order logic. In logic-based what-formalisms, the
model \textbf{c} of the world $\gamma$ around the physical computing
system $\alpha$ is usually characterized in terms of certain
set-theoretic structures, called $S$\emph{-structures}, where $S$
stands for the symbol alphabet of the formalism. Working out this
universal connection between logic and set theory early in the 20th
century was a milestone for mathematics. Logical inference (that which
happens, for instance, in Aristotle's syllogisms) became formalized
through inclusion relations between classes of $S$-structures. This
modern mathematical re-construction of logical inference reached its
final form in the work of Tarski \citep{Tarski36}.  Neither
$S$-structures themselves, nor inclusion relations between classes of
them, incorporate a reflection of time. These set-theoretic world
models can be intuitively seen as a collection of interrelated facts
embodied in structured sets, a static picture, not a dynamical
history. The ``computing'' that is formalized in
segment \textbf{a} and that physically happens in segment $\alpha$ is
cast as a process of logical reasoning about the structures modeled in
segment \textbf{c} through $S$-structures. In the 2370 year-old
perspective of logic, computing and reasoning are very much the same
thing. Logical reasoning proceeds by carrying out steps of logical
inference, like in syllogistic arguments. The \emph{premises} and the
\emph{conclusion} of each such an inference step are written down as
symbolic logic expressions. The conclusion that comes out of executing
one inference step become incorporated in the premises of the next
step, leading to \emph{inference chains}. Inference chains are
stepwise transformations of symbolic configurations, a fact which
establishes the connection to how-formalisms. A ``run'' (or
``execution'') of an algorithm can be understood as an inference chain
which leads from an initial premise to a final conclusion. The initial
premise is the \emph{input} to the algorithm and the final conclusion
is the \emph{output}. These two symbolic configurations lie in the
interface boundary \textbf{b}; they are the only items that are passed
to and fro between the reasoning system \textbf{a} and the environment
\textbf{c} that is being reasoned about. I return to the topic of
time. Logical inference is not \emph{temporal}. The relation between a
premise and a conclusion is not that the former comes first in time
and is temporally followed by the latter. Instead, the relation is
semantical-implicational: if the former is true, then the latter is
true too. Turing himself put this into plastic wording: \emph{It is
  always possible for the computer to break off from his work, to go
  away and forget all about it, and later to come back and go on with
  it.} \citep{Turing36}

By chaining, if the first input to an inference chain holds
true in the set-theoretic world model \textbf{c}, then so does the
ultimate output. This is the very same structure as of mathematical
proofs in general, which likewise proceed from initial premises or
axioms to the final claim in a series of argumentation steps. Hence,
the parlance in theoretical computer science to call certain
high-level algorithm formalisms \emph{inference engines} or
\emph{theorem provers}. But, if inference steps are natively
a-temporal, how then does time enter the picture?  This is a natural
question to ask, since after all human reasoning --- the original
inspiration for logic formalisms --- evolves in real physical
time. Any computational neuroscience that tries to understand human
logical reasoning in terms of physical brain dynamics must provide an
answer. But, in fact, this question is ignored in logic and computer
science textbooks. Here is how I see it. In a dramatic abstraction, parts
of reasoning in a physical brain can be regarded as a-temporal:
namely, if it is assumed that a reasoning human can \emph{store}
symbolic configurations in un-alterable, one could say platonically
immutable ways. An essential characteristic of symbols is that they
just stay identically the same once ``written''. Symbols are what does
not \emph{change}; they are exempted from time. These two facts are
two sides of the same coin: (i) that logic-based what-formalisms
capture a-temporal logical inference relations, which become
``update'' operations in how-formalisms, and (ii) that the native
substrate of how- and what formalisms are symbol alphabets. When such
how-formalisms are ``run'' on physical computing systems, the physical
hardware must necessarily provide for temporally unbounded, unaltering
\emph{memory} mechanisms where ``written'' symbols defy the mutations
of time. In physics terminology this means that digital hardware must
encorporate subsystems that have timescales far longer than the
use-time of the system. --- This is also the right context to
remark that by realizing logical theorem provers, digital computers
can carry out anything that can be found in a mathematical proof. To
the extent that some other, non-digital model of ``computing'' can be
mathematically formalized, it can be simulated by digital computers ---
except for the real-world temporal aspects of the simulatee. In this
sense, the digital computing paradigm is and will remain the master of
all others.

\emph{Take-home message \stepcounter{takehome}\arabic{takehome}:
  digital computing derives much of its superior powers from the very
  fact that electrical engineers and transistor developers have found
  ways to locally realize extremely large time constants, with time
  virtually coming to standstill.}

\textbf{Hierarchical structuring of formal constructs.} The formalisms
used in DC for modeling ``computing'' all admit to compose symbolic
configurations into more compounded ones, which then can be used as
building blocks in yet higher levels of \emph{compositionality}. That
human cognitive processing admits the creation of compositional
entities is regarded as constitutional for human intelligence by
proponents of classical AI and (Chomskian) linguistics.

Compositional hierarchies of formal constructs are present both in
how- and what-formalisms. I discuss the former first. Virtually all
programming languages (assembler languages possibly excepted) admit
the definition of compound configurations (from ``lists'' and
``arrays'' to ``modules'', ``objects'', ``scripts'') which bind
together more elementary symbolic configurations into larger
ones. Abstract models of ``computing'' systems likewise all have
provisions for defining compositional hierarchies, for instance by
joining symbols into nested sequences on a Turing machine tape, by
applying the lambda operator in lambda calculus, or by constructing
parse trees in grammar-based formalisms. Logic deductive theories, and
mathematical theories in general, create compound constructs by
hierarchies of formal \emph{definitions}. Compositional hierarchies of
symbolic configurations in how-models correspond to an execution
hierarchy of configuration update operations: ``executing'' or ``evaluating'' a
higher-level, defined construct means to execute a sequence of
lower-level update operations that happen inside the defined
construct. The higher a compound formal construct in a compositional
hierarchy, the larger the number of physical operations that are
needed to realize the execution of the formal construct on a
physical computing system. This multiplied physical effort leads to
longer processing real time for higher-level construct execution ---
unless some \emph{parallelization} scheme can delegate the required
physical operations to a multiplicity of physical sites that operate
simultaneously. Such time-space tradeoffs are an important topic in
theoretical compute science. It is generally desirable, but not easy,
to find formalisms and models of symbolic computing which lend
themselves to high degrees of parallelization. In contrast to current
models of symbolic computing, the human brain operates in an extremely
parallel fashion. We see a face together with its nose, mouth and
eyes. This is a lasting intellectual challenge for digital
computing theorizing, and has motivated the title ``Parallel
Distributed Processing'' for the bible
book \citep{RumelhartMcClelland85} that marks the start of neural
network research as we know it today.

The primary mechanism in DC what-formalisms (formal logics) to capture
the composition of cognitive entities employs nested \emph{functional
  expressions}, as for instance {\sf TakeoffWeight(AirbusA320,
  Fuelfill(FlightLH237), NumberPassengers(FlightLH237,
  January24\_2020))}. Such logic expressions \emph{denote} certain
structured sets contained in $S$-structures, and they can become
\emph{encoded} in (parts of) symbolic configurations in how-models.

\emph{Take-home message \stepcounter{takehome}\arabic{takehome}:
  Compositional formal structures are constitutive for DC models of
  computation. This makes DC how- and what-formalisms immediatly
  suited to capture the (widely but not uncontroversially claimed)
  compositionality of human cognitive entities.  The theoretical
  conveniences and formal powers afforded by symbolic compositionality
  are inseparable from hardly yet answered questions concerning the
  real-time, real-space realizability of formal configuration update
  operations.  }

\subsubsection{Probabilistic (sampling-based) formalisms and models}


I now turn to probabilistic models of computation, marked with P in
Figure \ref{figGrandSchema}. As stated earlier, in my view an
important criterion to call a physical system ``computing'' is that
its operations admit some kind of cognitive interpretation. For DC,
this cognitive aspect is rational logical inference. In probabilistic
``computing'' models the cognitive core aspect is the ability of
animals and humans to make probabilistic inferences. These are
formalized in probability formalisms through conditional probabilities
of the kind, ``if the sky is cloudy, then the probability of rain is
0.3''. Stating and evaluating such conditional probabilities requires
to have mental representations of \emph{probability distributions}. I
consider them the primary ``mental'' or ``cognitive'' objects in
probabilistic reasoning and formalisms, analog to symbolic
configurations in DC formalisms. Specifically, but not exclusively,
the need for evaluating conditional probabilities arises when agents
assess the chances that their actions will lead to the desired
outcome. This view of biological cognition is one of the leading
paradigms in cognitive science today and is referred to as
\emph{predictive brain} \citep{Clark13}, \emph{free energy model
  of cognition} \citep{Fristonetal10}, or \emph{Bayesian brain}
\citep{Tenenbaumetal06} hypothesis. In the areas of mathematics
and machine learning we find a wide diversity of worked-out
computational frameworks which formalize selected aspects of the
predictive brain perspective. They include models of computing with
probabilistic logic \citep{vonNeumann56}, logic-oriented accounts
of Bayesian statistics \citep{Jaynes95}, observable operator
models and predictive state representations of probability
distributions \citep{Jaeger00b, Littmanetal01}, Boltzmann machines
\citep{Ackleyetal85, HintonSalakhutdinov06}, reinforcement
learning approaches to modeling intelligent agents
\citep{Basyeetal95}, or the \emph{neural engineering framework}
\citep{Eliasmithetal12}. Here I restrict myself to formalisms and
models which use \emph{stochastic sampling} to represent and update
distributions. In methods for sampling-based processing of probability
distributions (SPPD for short), a probability distribution is not
mathematically represented by a closed formula. Instead, it is
approximately represented by a sample, that is a set of example points
``drawn'' from the distribution. Complex, high-dimensional
distributions cannot in general be represented by analytical
formulae. SPPD methods have become a major enabler for the
simulation-based study of complex systems in physics
\citep{Metropolisetal53}, for solving optimization problems, and
in some branches of machine learning. They are often referred to as
\emph{Monte Carlo} or \emph{Monte Carlo Markov Chain (MCMC)}
\citep{Neal93} or as \emph{particle swarm} methods
\citep{Dellaertetal99}. The requisite random sampling processes
still are mostly simulated on digital computers.  SPPD techniques have
however became also realized in non-digital physical computing
systems, namely in DNA computing \citep{vanNoortetal02} for
solving optimization and search problems, and more recently and with a
wider application range in analog spiking neuromorphic hardware
\citep{Indiverietal11, Haessigetal18, Moradietal18, Neckaretal19,
  Heetal19}. According to the \emph{neural sampling}
\citep{Buesingetal11, Pecevskietal11} view forwarded in
theoretical neuroscience, temporal or spatial collectives of neuronal
spike events can be interpreted (or used by the brain) as samples. Due
to this inviting analogy to biological brains and low-power
characteristics of analog spiking neurochips, research in such
hardware and corresponding ``algorithms'' is an energetically growing
field. But the main reason why I want to focus on sampling-based
versions of probabilistic models is that they open views on
``computing'' that differ from the DC views in interesting ways.

In order to preclude a false impression, I emphasize that there are
many other ways to represent distributions besides via sampling. For
instance, the procedural mechanics of certain (non-spiking,
non-sampling) artificial neural network models can be interpreted to
compute parameters of distributions, or probability distributions can
be approximately characterized and algorithmically processed through
variational calculus. Both methods are widely used in current
machine learning. Furthermore, firing sequences of neural spikes have
been interpreted to encode ``information'' in other ways than as
delivering stochastic samples. Precise firing time patterns of a
single neuron can be considered as carrying information in a
variety of ways \citep{Thorpeetal01, Izhikevich06, Deneve08a}.

Brushing over many specific differences between SPPD models, I will
now give an account of their common traits and place them
in the schematic of Figure \ref{figGrandSchema}. 

\textbf{Formalism interrelations.} I consider how-formalisms
first. There are two kinds of how-formalisms and how-models in SPPD,
depending on whether the targeted physical computing systems $\alpha$
are digital computers or not. In the first case, how-models are
symbolic algorithms which, when executed, generate sequences of
pseudo-random data points which accumulate over processing time into
samples. Such algorithms are called \emph{sampling algorithms} or just
\emph{samplers}. After down-compilation into low-level assembler
programs these sampling algorithms can be passed to physical digital
computers, crossing the modeling mirror. When the targeted hardware is
non-digital, how-models describe the physical mechanics of the
stochastic physical processes that are viewed as sampling
processes. For biological brains, theoretical neuroscience offers a
range of such models, from detailed physiological models of how spikes
are created in neurons to more abstract castings like
integrate-and-fire models of neurons or just Poisson spike trains. For
analog spiking neuromorphic microchips, the sample-generating,
physical stochastic processes on board of such microchips is modeled
with electronic engineering formalisms on the device and circuit
level \citep{Indiverietal11}.  I would think that there also exist
biochemical formalisms for modeling the reaction dynamics in DNA
computing microreactors, but I am unfamiliar with that field.

The situation for all the sampling algorithms that have been proposed
is markedly different from the situation in DC. In the DC domain,
how-formalisms and models can all be precisely related to each other
by mutual simulations. In SPPD I am not aware of a way how a stochastic
sampling algorithm A could ``simulate'' another such algorithm B. The
representation of a distribution by a sample is inherently
imprecise. The more data point examples are added to a sample, the
more precisely it captures the distribution. Different sampling
algorithms could possibly become related to each other by comparing
their statistical efficiency (how many sample points are needed for a
given accuracy level of representing a distribution). While this is an
important optimization objective in practical SPPD
designs \citep{Neal93}, I am not aware of a meta-theory or just attempts
to systematically set different SPPD algorithms in relation to each
other by comparing their statistical efficiency.

Turning to what-formalisms, these are the formalisms whose prime
mathematical objects are probability distributions, considered as
abstract objects independent of their concrete representation through
samples or other formats. What-models are written down in the
notation which students learn in probability or statistics
courses. They gain their expressive powers mostly from stating and
combining conditional probability relations. The analogy to the
picture I drew for DC is obvious. The if-then format of conditional
probability statements matches the if-then format of logical inference
rules. In fact, according to one influential view of Bayesian
statistics \citep{Jaynes95}, probabilistic what-formalisms can be
considered and worked out along the guidelines of formal logic. There
is however a noteworthy difference between DC and SPPD
what-formalisms. In the DC domain, different logics can be related to
each other through logic translations, yielding a (yet only partially
explored) ordering along an expressiveness scale. Nothing like this can
be found in probabilistic what-formalisms. There are no more or less
``expressive'' formalisms of probability theory. It makes some sense
to claim that there exists only a single probability what-formalism,
namely the one that is taught in probability theory
textbooks \citep{Bauer78}.

\emph{Take-home message \stepcounter{takehome}\arabic{takehome}: in
  both domains of DC and SPPD, formal models for computing systems
  came as how- or what-formalisms. Unlike in DC, how-formalisms in
  SPPD are not related to each other, and cannot be transformed into
  each other in an obvious way.}

\textbf{Intrinsic semantics.} SPPD what-formalisms are expressed using
the notation of mathematical probability theory. There are two major
epistemological schools of thinking about ``probability'', the
\emph{frequentist} (or \emph{objectivistic}) and the
\emph{subjectivist} (Bayesian) one. Their mathematical theorems almost
coincide in their surface format but are interpreted differently. In
frequentist interpretations ``probability'' means a physical property
of real-world systems, namely a system's propensity to deliver varying
measurement outcomes under repeated measurements, while in Bayesian
interpretations, ``probability'' means subjective degrees of
belief. The frequentist account of probability directly arises from
formal models of the physical environment (segment \textbf{c} in our
figure). These formal models of physical reality are a
\emph{probability spaces}, three-component mathematical structures
standardly written as $(\Omega, \mathcal{F}, P)$, where $\Omega$
(called ``universe'' or ``population'' among other namings) is a set
of \emph{elementary events} which can be intuitively understood as
locations in spacetime where measurements could possibly be made;
$\mathcal{F}$ is a certain set-theoretic structuring (a so-called
\emph{sigma-field}) imposed on $\Omega$; and $P$ assigns objective
probabilities to certain elements of this structuring. The formal
connection between such world models $(\Omega, \mathcal{F}, P)$ and
the descriptive what-formalisms that capture the probabilistic
cognizing about the world (segment \textbf{a}) is established by
\emph{random variables} which create segment \textbf{b}. I remark that
the term ``random variable'' is entirely misleading. Mathematically,
random variables are functions, not variables; and they are not
random, but deterministic. To make this clearer: a statement like
``Peter is male'' would be formalized in probability theory as {\sf
  Gender(Peter) = Male}, where {\sf Gender} is the random variable, a
function which deterministically returns the gender value from every
concrete person (formally: from every element $\omega \in
\Omega$). The randomness of random variables results not from that
they somehow return random values, but that random arguments are given
to them, following the probabilities prescribed by the item $P$ in the
world model $(\Omega, \mathcal{F}, P)$. The mathematics behind this is
involved and I allow myself at this point to recommend my probability
lecture notes \citep{Jaeger19c} where I attempt an intuitive, detailed
introduction to the basic concepts of probabilistic and statistical
modeling.

Summing up the
frequentist account of probability: The outside
world \textbf{c} is cast as a structured set $\Omega$ of observation
opportunities; observation apparatuses and procedures are abstracted
into random variables; measured values (from ticked gender
boxes in questionnaires to high-volume sensor data streams, filling
segment \textbf{b}) become the data points of samples which are one
way of formalizing probability distributions (segment \textbf{a}),
which in turn can be seen as the key formal correlates of concepts in
probabilistic accounts of cognition.

There are similarities and dissimilarities between the formalizations
of intrinsic semantics in DC versus SPPD. The what-formalisms in both
areas cast the physical world as highly structured sets
($S$-structures and sigma-fields) which represent, one might say, the
preshaped substance of the modeled piece of the world. Both DC and
SPPD what-formalisms are mathematically built around their respective
intrinsic semantics. However, the meaning of ``meaning'', that is, how
some symbol or formula (in segment \textbf{a}) relates to something in
the modeled world \textbf{c}, very much differs between DC and
SPPD. Frequentist probability theory includes  formal correspondents
of measurement or observation procedures and their values as
first-class citizens, namely random variables and the values that they
can take. Segment \textbf{b} is central in probability theory,
containing what is called \emph{sample spaces}. Probability theory universally separates the measurable
quality (like ``gender'', ``speed'') from the quantitative values that
this measurement categories can take, like ``male'' or ``100
km/h''. Qualities become cast as random variables, quantities as the
possible values of these variables. In contrast, the primary view
adopted by logic formalisms identifies quality with quantity. That
Peter is male would be formalized as {\sf Male(Peter)}. The symbol
{\sf Male} is a so-called \emph{predicate} symbol, and predications
(stating that certain objects have certain properties) are the most
elementary operations in first-order logic. It is however also
possible to express {\sf Gender(Peter) = Male} in logic. To do this
one has to introduce {\sf Gender} as a \emph{function} symbol. Still
there is an important difference. In the logical understanding of {\sf
  Gender(Peter) = Male}, all three items (the argument {\sf Peter},
the function {\sf Gender} and the value {\sf Male}) are contained in
the $S$-structure, that is they are both located inside the world
model in \textbf{c}. In a probabilistic interpretation of the same
English statement, {\sf Peter} is an element of the universe $\Omega$,
sitting in segment \textbf{c}; {\sf Male} is an observation value
which is mathematically placed outside of the world model
$(\Omega, \mathcal{F}, P)$ (I created the segment \textbf{b} to host
it), and {\sf Gender} links the two. This difference between logic and
probability modeling grows from deep historical roots.

\emph{Take-home message \stepcounter{takehome}\arabic{takehome}:
  Probability theory gives an account of how we \emph{observe} the
  world; logic is about how we \emph{reason about} an already observed
  world.}

There is one more similarity between logic and probability modeling
which will become central in Section \ref{secConcrete} for my
discussion of how MC theories may fit into the scheme of Figure
\ref{figGrandSchema}. Both SPPD and logic formalisms introduce
separate symbols for every observable quality of the elements of the
world-substance sets in segment \textbf{c}: random variables and
function symbols, respectively (like {\sf Gender}). An implicit
assumption behind this symbolic naming of qualities is that the named
quality is well-defined and \emph{remains the same} throughout the
time when the formalism unfolds in formal time (logic derivations or
formalizing sampling in \textbf{a}), and/or when this computational
process becomes physically instantiated (segment $\alpha$), and/or as
long as the real time evolves for the ``meant'' interpretation in the
real world environment $\gamma$ or its model \textbf{c}. This is a
highly nontrivial assumption and I will argue that it bars the way to
developing a formal theory of MC.

\textbf{Formal time} is knitted into SPPD formalisms and models in
intricate ways, often involving further theoretical elements
reflecting space and temporal hierarchies. In physical computing
systems (segment $\alpha$), real time is consumed on a fast timescale
to generate individual sample points, and on slower timescales to
accumulate sample points into samples. In the view of neural sampling
theories, where neural spikes (or possibly groupings of spikes)
correspond to sample points, the fast timescale is the one of single
spikes and the slow timescale the one of accumulating the effects of
single spikes by mechanisms of temporal integration, for instance via
building up neuron potentials or modulating synaptic
efficiencies. Formal models of spike event generation (in segment
\textbf{a}) can be expressed on different levels of granularity,
ranging from modeling the electrophysiology of spike generation with
differential equations to abstract representations of ``spike trains''
as Poisson processes. The accumulation of spike events into samples
requires extensions of these formalisms to include some kind of
temporal integration. --- When running sampling-based algorithms on
digital computers, the fast timescale reflects the runtime of program
subroutines to generate sample points; these subroutines contain
sub-subroutines to generate pseudo-random numbers which are normally
encapsulated as primitives (the {\sf rand} function) in programming
languages. Generated sample points are formally accumulated in lists
or arrays in program loops on the slow timescale of sample
build-up. --- This picture becomes more complex when sample points are
generated in parallel strands, leading, among others, to \emph{neural
  population} representations of samples in spiking neural network
models, or to \emph{Markov random field} models in non-neural
algorithms (though the latter are mostly used of purposes other than sampling). But no degree of parallelization can entirely cancel the
necessity for generating and accumulating sample points on respective
fast and slower timescales.

Because the creation of samples needs time (formal or real, segment
\textbf{a} or $\alpha$), the cognitive core items in SPPD, namely
distributions, are not defined for single moments of time or created
instantaneously. A sample-based representation (segment \textbf{a}) or
realization (in $\alpha$) of a distribution becomes the more precisely
defined the more sampling time is devoted to its
generation. Sample-based representations of distributions are
``smeared'' over time. If one adopts neural sampling views in
cognitive neuroscience, this has nontrivial consequences for the
notion of \emph{mental states} which I will not further pursue
here. The fact that samples develop over time leads to a spectrum of
ways of how SPPD models of computation can be used for practical
exploits, with different schemes for administering input. On one end
of the spectrum we find scenarios where the computing system is
``clamped'' to an unchanging input for the entire duration of the
computation. Formally this means that some random variables of the
model are frozen to fixed values. The sampling process is then run (in
formal abstraction \textbf{a} or physical realization $\alpha$) for an
extended period of time, allowing the sampling to grow the sample
large enough to decode from it the result with the desired degree of
accuracy (I skip the complications of initial transients and ensuring
ergodicity \citep{Neal93}).  Typical examples are classification
tasks (for instance, input is an image, desired result is a
probability distribution over possible classifications
\citep{HintonOsinderoTeh06}) or optimization tasks (the archetype
example: input is a roadmap, desired result is a round trip itinerary
through all cities which with high probability is the shortest one
\citep{Kirkpatricketal83}). On the other end of the spectrum, the
input is itself temporal, for instance a sensor data stream, and the
desired output is likewise temporal, for instance generated motor
commands in a robot or a brain, or an estimation of an agent's motion
relative to its visually perceived environment
\citep{Haessigetal18}, or speech-to-text recognition tasks.  This
is the generic situation in adaptive online signal processing and
control \citep{FarhangBoroujeny98} and in the study of
\emph{situated agents} \citep{SteelsBrook93}, which comprise
humans, animals, robots, software avatars or computer game
characters. Solving such tasks, where input patterns or output
patterns are themselves temporally evolving, requires sampling
mechanisms where the next generated sample point depends on the
history of previously generated ones. Important classes of formalisms
of this kind include spiking recurrent neural networks
\citep{Heetal19}, temporal restricted Boltzmann machines
\citep{Sutskeveretal09} and sampling-based instantiations of
dynamical Bayesian networks \citep{Murphy02a}. --- Intermediates
between these two extreme ends of the spectrum (single fixed input
versus nonstationary input streams) also occur, for instance when the
input is a sequence of fixed patterns which are clamped each for some
time before it is replaced by the next one.

\emph{Take-home message \stepcounter{takehome}\arabic{takehome}: The
  phenomenology of formal time in SPPD models is much richer than in
  DC. In DC, time reduces to discrete jumps from one well-defined,
  symbolic configuration the next, and these configurations are by
  themselves atemporal. In SPPD models, the core cognitively
  interpretable constructs, namely distributions, are themselves
  temporally defined through the sampling process. As a consequence,
  formal how-models must account for at least two timescales, and the
  formal interpretation of generated sample point sequences as
  distributions must account for nontrivial conditions like
  duration-precision tradeoffs or temporally overlapping realizations
  of different distributions. All of this is alien to the fundamental
  DC conception of executing algorithms.}

\textbf{Hierarchical structuring of formal constructs.} Human
cognitive processing admits --- or in some views
\citep{NewellSimon76}, is even constituted by --- the compounding,
or ``chunking'' \citep{Newell90}, of representational or
procedural mental states or mechanisms into larger compositional items
which then can again be composed again into even more comprehensive
items, giving rise to compositional hierarchies of representations,
mechanisms or processes. In DC this is accounted for by hierarchically
organized symbolic configurations (in how-formalisms) and nested
functional expressions in logic what-formalism. I can see three main
ways how distributions can become hierarchically organized.

First, some how-formalisms explicitly arrange their random variables
in layers, with ``low'' layers modeling the sensor data periphery and
``high'' layers modeling the cognitive interpretations of sensor
input. This is the case for many instantiations of Boltzmann machines
\citep{Ackleyetal85, HintonSalakhutdinov06}; furthermore, the
conditional dependency graph of random variables in Bayesian networks
and graphical models \citep{WainwrightJordan03} can be
hierarchically organized in reflection of a ``cognitive''
stratification. These are generic probabilistic information processing
formalisms. Task-specific layered spiking neural architectures, where
deep feedforward neural networks are being re-coded into spiking
neural substrates, have recently been receiving much interest, for
instance in visual object recognition tasks
\citep{Yousefzadehetal18}.  Furthermore, I am aware of two complex
information processing architectures which model complex human
cognitive dynamics on the basis of spiking neural network substrates,
namely Shastri's SHRUTI \citep{Shastri99a} series of connectionist
models human reasoning and language processing and Eliasmith's
almost-entire-human-brain models \citep{Eliasmithetal12}.

In these hierarchically or modularly structured systems, all random
variables in all levels or modules are, in principle, sampled from
with the same frequency. This is in agreement with the spiking
dynamics in biological brains, which likewise roughly has the same
timescale everywhere --- neurons in the visual cortex are not orders
of magnitude faster than neurons in the prefrontal cortex, although a
hierarchy of processing levels lies between them.  The ``cognitive''
constructs, namely distributions determined through through samples,
are formally the same throughout levels or modules. Higher-level
distributions are not made from, or composed of, lower-level ones.
This is different from the compositional hierarchies in DC models,
where higher-level symbolic configurations are made by binding
lower-level ones into compounds. In the SPPD models pinpointed above,
the assignment of being a ``higher'' or ``lower'' distribution is only
in the eye of the human inventor of the model.

Second, hierarchies of distributions canonically arise in
probabilistic models as conceived in Bayesian probability
\citep{Jaynes95, Jaeger19c} (note that ``Bayesian networks'' are
not Bayesian in this fundamental way of interpreting the nature of
probability; they got their name merely because they employ Bayes'
rule from elementary statistics). In Bayesian probability,
distributions become themselves distributed in
\emph{hyperdistributions}. In the original motivation of Bayesian
probability, these higher-level hyperdistributions reflect the
subjective prior beliefs of an intelligent agent about which
lower-level distributions are more or less plausible. Applying this
principle to modeling probabilistic cognitive systems one obtains
formalisms which are hierarchical in a substantial sense. The
relationship between a distribution and a hyperdistribution is
asymmetric. A hyperdistribution could be said to control, modulate or
``bias'' its lower-level children distributions. This gives rise to
cognitive processing models whose dynamics unfolds in an interplay of
bottom-up pathways (from sensor input to their cognitive
interpretations) and top-down pathways (cognitive expectations
modulating the perceptions through expectations). Prominent
representatives of such bidirectional cognitive processing systems are
Grossberg's Adaptive Resonance Theory models \citep{Grossberg13},
Friston's free-energy models of processing hierarchies in brains
\citep{Friston05} and Tenenbaum's models of human cognition
\citep{Tenenbaumetal06}. However, although Friston's writings
contain passing remarks on spiking neural dynamics, neither his nor
other's models of Bayesian cognitive architectures appear to have been
formulated or simulated (let alone physically realized) on the basis
of sampling processes. Instead, when these authors (as most other
neuro-cognitive modelers) want to capture temporal processing
phenomena, they use formalisms that abstract from sampling processes
by time-averaged neuronal firing rates governed by differential
equations (Grossberg, Friston). Or, when their focus is on the
a-temporal structure of hierarchies of (static) mental concepts, they
adopt the formalism of (Bayesian) probability theory. The Bayesian
distribution-hyperdistribution hierarchization should be analysable in
terms of sampling, at least if it is cognitively-biologically adequate
and if brains actually use spikes for sampling. This would be a
potentially rewarding mathematial project. Likely this has been done,
but I am not aware of it.

Third, more elementary (lower-dimensional) distributions can always
and precisely be mathematically combined into compound
(higher-dimensional) distributions by \emph{product}
operations. Conversely, complex distributions can sometimes more or
less approximately be \emph{factorized} into products of simpler
ones. Technically, this enables divide-and-conquer strategies for
efficiently computing probabilities in digital implementations of
graphical models \citep{HuangDarwiche94}.
Conceptually, this leads to insight into a specific kind of
compositional structure of complex distributions. Due to its practical
importance for probabilistic modeling with the aid of digital
computers, factorization algorithms for distributions are being widely
explored. Unfortunately, many real-world distributions cannot be
satisfactorily factorized. A generalization of products of
distributions is provided by \emph{tensor product} representations, a
standard operation in quantum mechanics \citep{Coecke12} which
also has been proposed as a paradigm for achieving analogs of symbolic
compositionality in distributed neural activation patterns
\citep{Smolensky90}. However, the computational efficiency gains
of factorizations are lost when the component distributions become
\emph{entangled} in tensor products. Like with Bayesian
hierarchization, I am not aware of sampling-based accounts of product
and tensor product operations, but again, this should be possible if
such product operations are cognitively adequate and if the brain
indeed exploits sampling for representing distributions.

\emph{Take-home message \stepcounter{takehome}\arabic{takehome}:
  Compositionality of cognitive representations, mechanisms and
  processes --- constitutive for symbolic computing --- has currently
  no clear analog in sampling-based models of computing. When sampling
  is used as the core process for enacting ``computing'' (as in
  Boltzmann machines or sampling-based evaluations of other graphical
  models), hierarchically organized interrelations between
  distributions exist only in the eye of the system designer but are
  not formally modeled on the level of interrelations between samples.
  To the extent that neural sampling is a valid view of information
  processing in biological brains, this means that either
  compositionality is not an inherent aspect of ``computing'', or it
  means that mathematicians still have to work out how hierarchies of
  distributions (like in distribution-hyperdistribution relations or
  factorizations of distributions) can be formally expressed through
  samples.  }

\subsubsection{Dynamical systems (ODE-based) formalisms and models}

I now turn to formalizations of ``computing'' that are rooted in
concepts and formalisms of dynamical systems theory, marked D in
Figure \ref{figGrandSchema}. Since almost a century, biological
systems --- neural and others --- have been studied in a line of
investigations which is referred to as \emph{general systems theory}
\citep{vonBertalanffy68} or \emph{cybernetics}
\citep{Wiener48}. This tradition co-evolved with the engineering
science of signal processing and control \citep{Wunsch85}. A
landmark in interpreting the human brain as a dynamical (self-)control
system is Ashby's classic \emph{Design for a Brain}
\citep{Ashby52}. In another co-evolving strand, neural dynamics
became modeled in a theoretical physics spirit, by isolating and
abstracting dynamical neural phenomena into systems of differential
equations, exemplified in the \emph{Hodgkin-Huxley} model of a neuron
\citep{HodgkinHuxley52}. Later, when the mathematical theory of
qualitative behavior of dynamical systems \citep{AbrahamShaw92}
had matured and in particular after \emph{chaos} and
\emph{self-organization} in dynamical systems became broadly studied,
dynamical systems modeling rose to a commonly accepted perspective in
cognitive psychology and cognitive science \citep{SmithThelen93a,
  GelderPort95}. Today the separations between these historical
traditions have almost dissolved. Mathematical tools from dynamical
systems theory are ubiquituously employed in modeling neural and
cognitive phenomena on all scales and abstraction levels, in a
diversity that defies a survey. Even when seen only from within
mathematics, dynamical systems theory is a highly diversified
field. Its formalisms range from finite-state switching systems to
field equations; time can be discrete or continuous; the \emph{state
  spaces} on which dynamics are described can be finite or infinite
sets, vector spaces, manifolds, function spaces, graphs or abstract
topological spaces, etc. Furthermore, this mathematical field has
overlaps with statistical physics, information theory, stochastic
processes, signal processing and control, game theory, quantum
mechanics and many more.

Here I will only consider models expressed with ordinary differential
equations (ODEs) which have the familiar look
$\dot{\mathbf{x}}(t) = f(\mathbf{x}(t), \ldots)$, where
$\mathbf{x} \in \mathbb{R}^n$ is an $n$-dimensional real-valued state
vector, $t \in \mathbb{R}$ captures continuous time and the three dots
``$\ldots$'' can optionally be filled with input signal terms
$\mathbf{u}(t)$ or \emph{control parameters} $\mathbf{a}$. The
function $f$ defines a \emph{vector field} on the \emph{state space}
$ \mathbb{R}$ which indicates the local direction and velocity of state
motion. Such ODE models are by far the most widely used and most
deeply studied kind of dynamical systems models; they are what
mathematics students find in their introductory
textbooks \citep{Strogatz94} (and physics students in theirs). The
mathematical theory of qualitative behavior in dynamical systems ---
attractors, bifurcations, chaos, modes, ``self-organization'', etc.\
--- has been first and foremostly been developed for
$\dot{\mathbf{x}}(t) = f(\mathbf{x}(t), \mathbf{a})$ system
models. The most pertinent reasons however why this choice of
mathematical substrate is appropriate in the context of this article
are that, (i) this is the most broadly used kind of formalism in the
modeling of continuous-time dynamics in biological and artificial
neural networks and cognitive processing; and (ii) it is the formalism
used by electrical engineers when they design analog ``neuromorphic''
circuits \citep{Indiverietal11}.

\textbf{Formalism interrelations.} When discussing DC and SPPD, I
found it useful to distinguish how- from what-formalisms. I will
follow this strategy again and separate the ODE formalisms and models
found in segments \textbf{a}, \textbf{b}, \textbf{c} in Figure
\ref{figGrandSchema} into how- and what-formalisms and models. The
how-formalism is the canonical textbook ODE formalism. Using its
notation, a researcher or engineer can write down how-models (in
segment \textbf{a}) which describe physical computing systems (in
$\alpha$). These models directly capture the real-time, metric
dynamics of continuous variables in computing systems --- voltages and
currents in analog microchips and neuronal circuits, ``activations''
of neural assemblies or concepts in models of neural cognitive
processes.

The what-formalisms describes how the state
trajectories $(\mathbf{x}(t))_{t \in \mathbb{R}}$ fold into the state
space $\mathbb{R}^n$, giving rise to the zoo of geometrical phenomena
which appear when state trajectories are traced over extended spans of
time and space in \emph{phase portraits}. Phase portraits are
structured and populated by ensembles of attracting or repelling or
saddle-node fixed points, oscillations, chaos, basins of attraction
etc. The fascinating geometrical worlds opened by studying qualitative
geometrical phenomena are most beautifully reavealed in the copious
picture-book \citep{AbrahamShaw92} of Abraham (mathematician) and Shaw
(artist) which I enthusiastically recommend as first reading for
novices in dynamical systems.

What-formalisms for dynamical systems may look superficially analog to
the what-formalisms that I discussed in logic-based and probabilistic
modeling. However, in those two modeling worlds the what-formalisms
have been worked out into canonical formats (formal logic and probability
theory), and the relationships between how- and what-formalisms are
transparently defined. This is not the case in ODE modeling, let alone
in dynamical systems modeling in general. There is no canonical,
complete mathematical formalism to comprehensively describe the
world of geometric wonders that arise in ODE systems. Invariably,
mathematicians use plain English besides formulas to
describe how attractors etc.\ become geometrically or dynamically
related to each other in phase portraits. The very notion of a phase
portrait, ubiquitously used in mathematical texts, is itself
not formally defined.  New kinds of qualitative phenomena are
continually being discovered. Most insights into the geometry of
dynamics that are today available have been discovered, defined and
studied, by mathematicians and theoretical physicists, for
\emph{autonomous} dynamical systems only. These are systems whose
equations $\dot{\mathbf{x}}(t) = f(\mathbf{x}(t), \mathbf{a})$ have no
input term. The study of qualitative behavior in input-driven systems,
which have an addditional input term $\mathbf{u}(t)$ in their ODEs, is in its
infancy \citep{KloedenRasmussen11, ManjunathJaeger13a}. Current
mathematical theory offers only painfully limited ways of inferring
from a given how-model
$\dot{\mathbf{x}}(t) = f(\mathbf{x}(t), \mathbf{a})$ or
$\dot{\mathbf{x}}(t) = f(\mathbf{x}(t), \mathbf{u}(t), \mathbf{a})$
which qualitative phenomena emerge from it. This has to be worked out
on a case-by-case basis. Analytical understanding is often impossible
to achieve, and numerical, intuition-guided simulations on
low-dimensional subspace projections afford the only window of
insight.

It might be well the case that a comprehensive what-formalism, and
what-theory expressed on its basis, is principally impossible for ODE
systems. Just like natural scientists will forever continue to
discover new qualitative phenomena in nature, and just like biological
evolution incessantly finds and exploits new qualitative phenomena on
a ``keep \emph{any}thing that works'' basis, mathematicians may be in
for a principally open-ended discovery journey. This open-endedness
may be intrinsic in an ill-definedness of what \emph{qualitative}
means.  If this is so, we are facing a serious problem with regards to
modal computing:

\begin{enumerate}
\item \textbf{if} a physical system should be called ``computing'' only
  if its operations admit some kind of interpretation in cognitive terms,
\item \textbf{and if} it is the qualitative phenomena in ODE systems
  which can be subjects of ``cognitive'' interpretations, rather than
  the quantitative ODE mechanics,
\item \textbf{and if} the leading idea for MC is to exploit \emph{any}
  physical phenomenon that is useful for ``computing'',
\item \textbf{and if} no comprehensive, closed what-theory is possible
  due to a principally open-ended richness of qualitative phenomena,
\item \textbf{then} no complete, closed theory of MC on the basis of
  ODE dynamical systems modeling is possible.
\end{enumerate}

This is disconcerting. Physicists rely heavily on ODEs and other kinds
of differential equations to model physical systems, and so do
electrical engineers when they design circuits for analog
microchips. ODE formalisms seem thus a natural choice for developing a
theory for MC, which should be able to connect to  the physics of computing
systems. I see four options to deal with this roadblock:

\begin{enumerate}
\item Accept
that a closed, complete theory is not possible and embark on MC theory
building in the spirit of an open-ended discovery adventure.
\item Build an MC theory around an apriori restricted set of qualitiative
phenomena. This is what DC does in relying, deep down, only on the
phenomenon of bistability which gives the 0-1 bits DC is made of. This
is also what we see in most neural network models in comutational
neuroscience and machine learning. Three examples: Freeman's account
of neural pattern representation and recognition on the basis of
chaotic attractors \citep{YaoFreeman90}; Rabinovich's models of
cognitive sequencing of concept activations along heteroclinic
trajectories between saddle nodes \citep{Rabinovich08}; or my own
proposal for organizing and processing neural memory with certain
nonlinear filters called ``conceptors'' \citep{Jaeger17}. The obvious
drawback is a limitation of perspective which will miss the vast majority of
useful phenomena.
\item Not adopt a dynamical systems oriented
perspective. This can be utterly successful as witnessed by DC, but I
am afraid it bars the way to exploiting all that physics can
offer.
\item Find a new foundational mathematical formalism for
dynamical systems which makes qualities, not quantities, the objects
of dynamical change. This is what I find the most promising route, and
I will put forth initial ideas in the final section of this article.
\end{enumerate}

I return from this excursion into methodological
questions and take a closer look at interrelations between ODE
models. To avoid misunderstandings I start with a clarification. ODE
models are used both in connection with digital and with analog
computing hardware. This is done in very different ways. The use-case
with digital computers: Researchers who model some dynamical
real-world system of their respective discipline --- physicists
modeling a mass on a spring, biologists modeling predator-prey
interactions --- write down ODE equations and then simulate them on
their digital machine. This use-case falls into the DC part of this
survey. ODEs are here a what-formalism and the written-down equations
a what-model, which in order to become executable has to be ``coded''
in a how-formalism, that is, a suitable programming language. In contrast,
in connection with analog hardware, ODEs serve not to
simulate some other dynamical system on the used hardware, but
to model the physical computing system itself. I consider here
only this second use-case.

In DC, different how-formalisms (programming languages, abstract
models of ``computation'') can be be encoded into each other. For SPPD
how-formalisms (formalisms to specify samplers, neural spike
generation mechanisms or DNA sniplet formation), mutual encodings have
not been studied as far as I can see. In the dynamical systems
modeling domain, if we restrict the discussion to ODE modeling, there
is only a single how-formalism, namely the ODE formalism, and the
question of formalism interrelations is moot. If we would include
other dynamical systems formalisms into our discussion, numerous
mutual encoding relations can be found. In particular,
continuous-time, continuous-space formalisms (ODEs, partial
differential equations, field equations) can be arbitrarily well
approximated by discrete-time, discrete-space formalisms (like
numerical ODE solvers, cellular automata, finite-element
formalisms). Unlike in DC, where such formalism translations are
precise and transparent and can be easily established in student
homework exercises, these continuous-to-discrete encodings in the D
domain are approximate, finding them is not trivial and analysing the
approximation quality can be very difficult.

\emph{Take-home message \stepcounter{takehome}\arabic{takehome}: In
  dynamical systems oriented modeling of computing systems (restricted
  here to ODE modeling), the single how-formalism is the textbook ODE
  formalism. In computing sytems modeled by ODEs, an unbounded
  plethora of geometrical, \emph{qualitative} phenomena may
  arise. They are candidate formal objects to be interpreted in
  what-models as correlates of ``cognitive'' phenomena. A few of them,
  in particular fixed points, attractors, bifurcations and modes, are
  already widely being perceived in the cognitive and neurosciences. }

\textbf{Intrinsic semantics.} In L and P modeling there are canonical
ways how the environment ($\gamma$ in Figure \ref{figGrandSchema}) is
mathematically modeled (in segment \textbf{c}), namely by the
set-theoretic constructs of $S$-structures and probability spaces. In
dynamical systems oriented modeling no canonical view on how to model
the environment exists. The question is rarely asked and I dare say
under-researched. There are historical reasons for this semantic
almost-blindness. Starting with Newton, dynamical systems modeling has
for a long time been the homeground of physics. Experimental
physicists try hard to \emph{isolate} their system of interest from
the environment. This made them use a kind of ODE which are, in
mathematical terminology, \emph{autonomous}, that is, their equations
$\dot{\mathbf{x}} = f(\mathbf{x}, \mathbf{a})$ have no input term. The
historical development of dynamical systems mathematics was by and
large confined to autonomous systems. Starting, say, in the 1960ies,
there was explosive mathematical progress in perceiving, analysing and
(importantly) visualising \citep{Peitgen86} qualitative phenomena like
attractors or chaos. Other disciplines besides physics and even the
general public became fascinated by dynamical systems and began to
interpret their respective objects of study as qualitative dynamical
phenomena. In the cognitive sciences this became a established
perspective around the year 1990 \citep{Schoeneretal86,
  PortGelder95}. However, the available mathematical tools and
metaphors were rooted in, and confined to, autonomous input-free
systems. This barred the way to develop dedicatedly semantic accounts
of how the modeled systems interact with their environment. In neuro-
or cognitive modeling, qualitative phenomena inside the modeled neural
or cognitive system were not related to its outside environment, but
were mapped to system-internal cognitive constructs and processes. An
important theme was (and is) how symbols and ``concepts'' can be
interpreted by qualitative dynamical phenomena --- the neuro-symbolic
integration problem \citep{Besoldetal17}.

In my opinion, in order to work out a genuine intrinsic semantics for
dynamical systems models of ``computing'' systems, one would have to
move from autonomous system models like
$\dot{\mathbf{x}} = f(\mathbf{x}, \mathbf{a})$ to input-driven ones
like $\dot{\mathbf{x}} = f(\mathbf{x}, \mathbf{u}, \mathbf{a})$. But,
as I remarked earlier, mathematical theory development for qualitative
phenomena in such \emph{non-autonomous} systems is far less developed
than for autonomous ones \citep{KloedenPoetzsche13}. Even
clarifying the notion of an attractor --- the most focussed dynamical
phenomenon in current dynamical systems oriented cognitive modeling
--- is loaden with mathematical intricacies \citep{Poetzsche11,
  ManjunathJaeger13a}. It will take a while before this venue can be
widely explored outside an inner circle of specialized mathematicians.

An easier alternative to embarking on non-autonomous dynamical systems
is to cast the entire system comprised of both the ``computing'' agent
and its environment as a single autonomous system. The computing
system is then seen as a \emph{subsystem}. This view has been adopted
in two classical dynamical systems models of biological agents, albeit
not based on ODEs but on discrete-time, finite-state cellular
automata. The first is von Neumann's model of self-replicating
automata \citep{vonNeumann66} (which also marks the invention of
cellular automata in the first place), the second is the model of
\emph{autopoietic} living systems by Varela and Maturana
\citep{Varelaetal74}. In both models, the ``computing'', rather:
``living'' system is modeled as a delimited, moving, shape-changing
and growing area within a cellular space-time grid. The theory of
autopoietic systems is, in a sense, explicitly anti-semantic. What
happens inside such a system is not reflecting, or representing,
outside givens: \emph{``In this sense we will always find that one
  cannot understand the phenomenon of cognition as if there were
  'facts' and objects 'out there', which one only would have to fetch
  and put into the head. [...] The experience of every thing 'out
  there' becomes configured by the human structure in a specific
  way...''} (my translation from a German translation
\citep{MaturanaVarelaEnglish84} of Maturana and Varela's book
\emph{El \'{a}rbol del concocimiento}). Instead, an autopoietic system
constructs its own internal world, which is shaped and connected to
the outside only because this organization has to meet requirements of
self-stabilization and survival. Maturana and Varela have coined the
term \emph{structural coupling} for this principle
\citep{MaturanaVarelaEnglish84}. Their biologically motivated
theory has given rise to schools of epistemology called (radical)
\emph{constructivism} \citep{Schmidt87} and \emph{enactivism}
\citep{WilsonFoglia17}. As far as I am aware, a mathematical
formalization of structural coupling in terms of qualitative phenomena
in dynamical systems has not been attempted.

ODE-based models of ``computing'' agents interacting with their
environment are naturally designed as two coupled ODE subsystems. Call
the system state vectors of the agent and the environment
$\mathbf{x}_a$ and $\mathbf{x}_e$, respectively. The agent can
perceive the environment through \emph{observations}
$\mathcal{O}(\mathbf{x}_e)$, and the environment is influenced by
actions, or \emph{controls} $\mathcal{C}(\mathbf{x}_a)$. This leads to
coupled equations
$\dot{\mathbf{x}_a} = f_a(\mathbf{x}_a, \mathcal{O}(\mathbf{x}_e)),
\dot{\mathbf{x}_e} = f_e(\mathbf{x}_e, \mathcal{C}(\mathbf{x}_a))$
which fit precisely into the segments \textbf{a} and \textbf{c} of
Figure \ref{figGrandSchema}, with the observations and controls placed
in segment \textbf{b}. Such models are often used in nonlinear systems
and control engineering (though here the match with Figure
\ref{figGrandSchema} is incomplete since the ``controller'' subsystem
$\mathbf{x}_a$ also receives a \emph{control target} input from
outside the modeled ``plant'' $\mathbf{x}_e$). Such models have also
been discussed in some works in the theory of intelligent agents, in
particular Ashby's classic model of the brain \citep{Ashby52} and,
much later, Beer's pledges to model cognitive-biological agents in
terms of continuous dynamical systems rather than within the
conceptual framework of symbolic computing \citep{Beer95b}. The
fit into Figure \ref{figGrandSchema} is here precise because this line
of research aims at modeling \emph{autonomous} agents which do not
receive external ``target'' signals but instead generate their goals
internally. In control engineering and also in the autonomous agent
modeling lines, the focus of modeling lies on understanding conditions
for dynamical \emph{stability}. Qualitative, cognitively interpretable
phenomena arising in the agent subsystem
$\dot{\mathbf{x}_a} = f_a(\mathbf{x}_a, \mathcal{O}(\mathbf{x}_e))$
are not typically considered in control engineering. In
low-dimensional simulation experiments \citep{Beer95b}, Beer
discusses how geometric structures in the phase portrait of the agent
$\mathbf{x}_a$ relate to geometric information obtained from the phase
portrait of the environment $\mathbf{x}_e$. Beer's aim is to
demonstrate the general usefulness of dynamical systems formalism for
studying autonomous behavior. As far as I can see, research aiming at
a systematic mathematical modeling of intrinsic semantic relationships
between agent $\mathbf{x}_a$ and environment $\mathbf{x}_e$ subsystems
has not been attempted yet. It would be most rewarding to develop
mathematical approaches to understand qualitative phenomena which are
simultaneously emerging in two subsytems $\mathbf{x}_a$ and
$\mathbf{x}_e$ which interact with each other through observation or
control signals or, more generally, through shared variables.

\emph{Take-home message \stepcounter{takehome}\arabic{takehome}: The
  study of how cognitively interpretable, qualitative phenomena
  arising in the internal dynamics of an agent relate to dynamical
  models of its environment has barely begun for ODE
  formalisms. Appropriate ways of achieving this may have to wait for
  a further development of non-autonomous dynamical systems
  theory. Existing dynamical models of agent-environment interaction
  use the same formalism for both. This is different from
  logic-symbolic and probabilistic modeling where the external
  environment is mathematically cast differently from the
  ``computing'' system. In existing agent-environment dynamical
  models, both parties are modeled on the basis of the same carrier
  substrate, namely dynamical (sub)systems described with the same
  formalism. }

\textbf{Formal time.} The formal time model is hidden in the dot in
$\dot{\mathbf{x}}$, which is the common shorthand for the derivative
$\delta \, \mathbf{x} / \delta \, t$ with respect to time $t$. This
real-valued time $t \in \mathbb{R}$ is considered in physics,
neuroscience and engineering as capturing the continuous arrow of
physical time, one of the reasons why ODE formalisms are so natural
for physicists and engineers. The connection to physical time is
obvious, direct and convincing for natural scientists and
engineers. This is fundamentally different from the formal time in DC
modeling, which is logical-inferential in its origin and becomes
mapped to the physical time of physical computers in essentially
arbitrary scalings (faster on faster computers). The fact that a
formal ``1 sec'' segment of $t$ really becomes one physical second
when ODE circuit models are realized in analog electronic hardware has
important and not easy to deal with consequences for analog computing
practice. First, the designer of analog circuits must match time
constants in his/her formal ODE models to the physical time on board
of the microchip. Second, assigning different time constants to
different system variables is an utterly delicate affair, because
slightly different settings of these time constants may lead to
qualitatively different system dynamics. Third, the physical input
signals to (non-autonomous) ODE systems evolve in the same physical
time as the computing system. Thus, analog computing systems must be
timescale-matched to their input-delivering environment.  On the
positive side, this makes it natural to design analog computing
systems that are physically embedded in their environment through
continuous input- and output signals. Digital computing faces
deep-rooted difficulties with continuous real-time processing and the
common solution there is to build or buy machines whose digital clock
frequency is so fast that the inevitable delays of ``processing'' the
current input value become non-disruptive for the task at hand. It is
however technically difficult to construct analog hardware that is
both small-sized and slow. In analog electronics, small-sizedness
conflicts with slowness because slow time constants spell out into
large capacitors or very low voltages, both of which quickly hit
limits. In our own work with analog neuromorphic microchips we found
this the most challenging obstacle when we set out to implement an RNN
for a task of online heartbeat classification from real-time ECG
signals: these signals were too slow for our microchip and we had to
find ways to exploit collective phenomena in the RNN to create slow
derived system variables \citep{Heetal19}.

These difficulties render the practice of designing and using analog
(especially neuromorphic) computers in ``situated'' online processing
tasks markedly different from how digital machinery is used. One
cannot use the same machine and neural network model for fast and for
slow tasks. The formal model timescales must match the physical
machine's which in turn must match the ones of the task. In DC, the
only limit regarding time in situated online processing scenarios is
given by the digital clock frequency. Provided that adequate
``real-time'' operating systems and programs are available, any task
with slower timescales can be solved with digital machines by
buffering variable values for waiting times as long as the slow task
timescale demands. Furthermore, the qualitative behavior of analog
circuitry and their (RNN) how-models depends sensitively on the
settings of numerical parameters in the models, especially (but not
exclusively) on time constants. A similar fine-tuned dependency on
physical and model numerical parameters is no issue in DC. On the plus
side: analog computing systems that are directly embedded in their
task environment through real-time input and output signals can, in
principle and like biological brains, cope with broad-band signal
interfacing with virtually zero delay (up to speed of electric signal
propagation limits).

\emph{Take-home message \stepcounter{takehome}\arabic{takehome}: In
  ODE-based how-models the timescales of the model must match the
  physical timescales of the input / output signals if the physical
  computing system is used in online tasks. This is similar to the
  external operating conditions and internal mechanisms of biological
  brains, but very different from digital computing.  }

The \textbf{hierarchical structuring of formal constructs} is
intimately connected with timescales in ODE modeling. When researchers
in cognitive neuroscience, robotics and autonomous agents or machine
learning conceive of their respective intelligent agent architectures
as dynamical systems, they almost by reflex assign \emph{fast
  timescales} to subprocesses that operate close to the sensory-motor
interface boundary (segments \textbf{b}, $\beta$), and the assign
\emph{slow timescales} to subprocesses operating at ``higher''
cognitive levels. Multiple timescale dynamics are of general interest
for \emph{complex systems} modeling in all natural sciences.  A rich
body of mathematical theory has grown \citep{Kuehn15}. When
writing down ODEs, slow versus fast timescales can be imposed on
system variables by multiplying \emph{time constants} into their
differential equation. Also, in ODE systems all of whose variables
have the same (fast) time constant, new descriptive variables can be
discovered through mathematical analyses which reflect slowly changing
collective characteristics of the underlying system. This can be done
in many ways. For instance, complex chaotic attractors can be
described in terms of their make-up from connected \emph{lobes}, which
can be discerned from each other geometrically (by approximating them
with periodic attractors) or by registering dwelling times (the
dynamics stays for some time in one lobe before it moves to another)
or by the possibility to stabilize the lobes
\citep{BabloyantzLourenco94}.  In computational neuroscience and
machine learning this is explicitly and effectively exploited in
\emph{slow feature analysis} \citep{WiskottSejnowski02} where
slowness of derived variables makes them interpretable as cognitive
representations of object concepts.  In physical systems, the
emergence of multiple timescales is often connected to mechanisms that
cannot be captured with ODEs, in particular delays in system-internal
signal propagation and spatial extension of systems.

Most qualitative phenomena arising in ODE systems are described via
static geometric-topological conditions in phase portraits. Time is by
and large factored out in the definitions of phenomena like fixed
points, attractors or bifurcations. The ``speed'' of state variable
evolution is irrelevant for the mathematical definition of these
constructs. Time enters these definition at best in the form of
asymptotic (infinite duration) characterizations of
convergence. With maybe the exception of a spatial hierarchical
structure described for chaotic attractors or certain structures in
bifurcation cascades, these ``timeless'' geometry-defined phenomena
offer no hooks for hierarchicity. In my opinion, timescale hierarchies
are key for finding hierarchicity in dynamical systems. Compared to
the rich and advanced state of the art in characterizing
geometry-defined phenomena, the study of timescale-induced phenomenal
structuring is in an early stage. There is, to my knowledge, no
established terminology for discussing this theme. I used the word
``mode'' earlier in this article to denote the condition that a
nonstationary or non-autonomous dynamical system transits through a
sequence of hierarchically nested periods where within each period it
exhibits dynamical or statistical phenomena that are characteristic
for this period. The word ``mode'' is suggestive and variously used in
the literature, but it does not have a canonical definition. I use it
in the (admittedly vague) sense that the values taken by slower
variables characteristically modulate the behavior of the respective
next-faster variables.  How this modulation is discerned or quantified
is up to the mathematician or physicist who tries to understand
``what'' is going on. A mode could be specified or quantified, for
instance, by characteristic frequency mixtures, the visited region in
state space, specific signal shapes or begin- and end-marking events.

Mathematical and applied research on multiple timescale dynamics in
ODE models is so rich that a comprehensive treatment cannot be
attempted here.  I single out one kind of system model which offers
relevant insights for understanding ODE systems as
``computing''. These are recurrent neural network (RNN) models where a
stack of neuronal ``layers'' corresponds to a sequence of increasingly
slower timescales, starting from a ``low'' fast sensor-motor interface
layer to ``higher'' layers whose increasingly slower neuronal state
variables are interpreted as representing increasingly compounded
``features'' or ``concepts''. Examples are RNN based robot control
systems where these timescale differentiations are predesigned into
the neuronal hierarchy through time constants
\citep{YamashitaTani08}; or a hierarchical RNN model for the
unsupervised discovery learning of increasingly slower components in
multiple-timescale input signals, where the different timescales are
predesigned by way of using faster or slower ``leaking'' in the used
neuron model \citep{Jaeger07b}. In neuroscience, the idea that
biological brains operate at fast timescales close to their
sensor-motor periphery, and at increasingly slower timescales as one
moves ``upward'' in the anatomical hierarchy toward forebrain
structures, has been advanced on the basis of theoretical argument
corroborated by simulation studies \citep{Kiebeletal08}, and it
also has been confirmed by physiological measurement
\citep{Murrayetal14}. In machine learning architectures the formal
neurons in higher layers are made slow by adorning their ODEs with
slower time constants --- ``higher'' neurons are slower neurons. With
regards to biological brains, it is still unclear to which extent (and
on the basis of what physiological mechanisms) individual neurons in
different brain regions are slower or faster; or how or to what extent
different timescales of \emph{neural integration} of information
arises from collective mechanisms
\citep{GoldmanCompteWang07}. Even when all neurons across the
hierarchy are individually equally fast, slowness in higher layers can
arise as a property of derived variables like spiketrain
autocorrelation timescales \citep{Murrayetal14} or collective
variables like local field potentials \citep{Kiebeletal08}.

The distinction between how- and what-models becomes a little blurred
in hierarchical RNN models. They can be regarded as how-models
inasmuch as they are written down as ODE systems which can become
realized by electrical engineers in analog neuromorphic circuitry. On
the other hand one might consider timescales a geometric, qualitative
property that admits a cognitive interpretation. Then, according to my
proposed informal definition, these models appear as
what-models. This view is immanent in the intuitions about physical
intrinsic semantics held by those cognitive modelers who posit that
brain-internal variables represent objects or situations in the
external environment, as revealed in this quote: \emph{``Many aspects
  of brain function can be understood in terms of a hierarchy of
  temporal scales at which representations of the environment
  evolve. The lowest level of this hierarchy corresponds to fast
  fluctuations associated with sensory processing, whereas the highest
  levels encode slow contextual changes in the environment, under
  which faster representations unfold.''} \citep{Kiebeletal08}.

The low-to-high layering in hierarchical RNN architectures can be
\emph{feedforward} (lower layers feed only to higher layers but not
vice versa) or \emph{bidirectional} (with both ``bottom-up'' and
``top-down'' couplings between the layers. Biological brains are
eminently bidirectional, as are most formal models in computational
neuroscience and machine learning. I will only consider bidirectional
models here. With regards to timescales there are noteworthy
similarities and dissimilarities between bidirectional hierarchical
RNN models and digital computer programs. Large computer programs are
written by professional programmers in a hierarchically organized
fashion, with higher-level modules (subprograms, scripts, functions,
objects) calling lower-level ones as subroutines. Such programs are
bidirectional too: the higher-level module ``calls'' the lower-level
one and typically sends down initialization parameters; and
conversely, the lower-level subroutines send the results of their
computation upwards to the calling module.  As I remarked earlier,
this automatically leads to a runtime hierarchy: executing
higher-level modules necessarily takes longer than executing the
lower-level because the latter are called inside the higher-level
one. The resulting timescale hierarchy could be called a waiting-time
hierarchy: the symbolic configuration update step assigned to the
higher-level module is pending in an indeterminate status as long as
the called lower-level subroutines are being executed. In contrast, in
bidirectional RNN systems, all neurons on all levels are active
simultaneously, and the top-down and bottom-up information flows are
continually streaming without need (or opportunity) for waiting. This
is yet another reason why brains cannot easily be likened to symbolic
computing machines, and why timescales are a key issue when one wants
to develop conceptualizations of ``computing'' that extend beyond the
DC paradigm and include brains.

For mathematical analyses of top-down influences from higher to lower
layers it is decisive \emph{how much} slower the higher layer
develops. If the \emph{timescale separation} is very large (say two
orders of magnitude or more), the slow evolution speed in the higher
layer can be approximately regarded as standing still compared to the
fast lower layer. It is then possible and common practice to consider
the variables fed to the lower from the higher layer as constant
control parameters in the equations governing the lower layer, and
different processing ``modes'' of the lower layer can be separately
studied by considering different static settings of these control
parameters. If the timescale separation is not very large, analyses
become difficult.  In RNN and cognitive dynamics research, one access
route is to trace the emergence and decay of transient qualitative
phenomena in the lower layer as being guided by the attractor
structure that would be defined if the timescales were widely
separated \citep{Werneckeetal18}.

Strictly layered architectures are popular in machine learning, AI and
control engineering \citep{Albus93}. When they are bidirectional, they
probe the limits of today's mathematical analyses. Biological
brains \citep{FellemanVanEssen91} and complex practical control
architectures \citep{Thrunetal06} have a modular structure which is more
complex than a linear ordering of processing layers, with many lateral
and diagonal processing pathways. Furthermore, it may happen that a
variable that is fast at some time may turn into a slow one later and
vice versa. Analysing, or merely characterizing such timescale
meta-dynamics is beyond the reach of today's mathematics.

\emph{Take-home message \stepcounter{takehome}\arabic{takehome}:
  Timescale hierarchies, possibly structurally reflected in layered
  architecture designs, appear to be an important ingredient of
  dynamical systems models to qualify them as ``computing''. Different
  timescales in (neuromorphic) analog hardware and their mathematical
  models can arise from a variety of formal and physical effects, for
  which we still lack a comprehensive understanding. The formal nature
  of multiple timescales differs fundamentally between digital
  computers (waiting time hierarchies) and parallel analog systems
  (continually interpenetrating bottom-up and top-down streams of
  information with different time constants).}

\subsection{Section summary}\label{secSummary}

The purpose of this section was to help myself and the reader to
understand better what we may mean when we speak of a ``theory of
computing''. This is a fairly well demarcated concept in symbolic /
digital computing --- there is a choice of university textbooks for
``Theoretical Computer Science'' which by and large all present the
same canonical material. When today we speak of ``computing'', our
understanding of this word is pre-shaped by the paradigm of symbolic
computing to a large extent.  But brains work differently from digital
computers, and so will future non-digital hardware systems,
neuromorphic or otherwise. I proposed the name ``modal computing''
(MC) as an umbrella term for any approach to engineer physical systems
which aims at exploiting nonlinear ``modal'' physical phenomena for
``computing'' --- just as cleverly, opportunistically, and
resource-efficiently as biological brain evolution did. In order to
establish MC as an engineering science --- a faraway goal --- new
theories have to be developed which formally codify a
conceptualization of ``computing'' that is as different from the DC
paradigm as a brain is from a desktop computer, and maybe even more
different. This is a voyage into the unknown. But it is not a voyage
blindfolded. A wealth of relevant ideas, experimental designs, and
formal theories has already accumulated in a diversity of disciplines
and historical lines of thinking. But its richness and diversity is
also confusing. In this section I attempted to find compass
coordinates to help navigating in this intellectual heritage.

An exhaustive exploration of all existing relevant insight is beyond
anybody's means. I had to constrain my exploration by a number of
decisions which reflect my limited knowledge and personal views:

\begin{itemize}
\item From the outset I declare a set of four conditions which I deem
  necessary for a physical system to be ``computing'', namely that it
  operates in time, that it is open to input or output, that some
  aspects of it must be interpretable in cognitive terms, and that it
  must admit a semantic interpretation.
\item I restrict myself to investigate formal mathematical theories
  only, which means I ignored a bounty of philosophical and 
  empirical insight.
\item As a coarse scheme for organizing my exploration I adopted a
  simple-minded division between \emph{physical} computing
  systems, their environment and the interface boundary between the
  two on one side, and their respective \emph{formal} models on the
  other side of what I called the ``modeling mirror'' (Figure
  \ref{figGrandSchema}).
\item I divided formal models of computing systems into two
  classes. How-models capture the ``mechanics'' of computing
  processes, and what-models give accounts of the computing processes
  in cognitively interpretable categories which admit semantic
  mappings to the system's environment.
\item I limited my selection of modeling approaches to the ones whose
  underlying host mathematics is logic, probability theory, or
  dynamical systems theory (L, P and D, Figure
  \ref{figZoomIn}). Within the P and D domains I further restricted my
  coverage to sampling-based models and ODE based models,
  respectively.
\end{itemize}

In each L, P, D domain I discussed four themes: mutual transformation
/ translation interrelationships between formalisms; formal semantics;
how time is formalized; and how constructs within a formalism are
hierarchically organized. This choice is undoubtedly influenced by my
lifelong exposure to digital computing theory.  Except maybe for time,
these themes are central theorizing coordinates for digital computer
science. All the non-digital information processing models (in the P
and D domains) that I visited reveal that these themes can be worked
out in significantly different ways than we know it from DC. What we
see in those other domains can be called ``computing'' on the grounds
that

\begin{itemize}
\item the respective authors call it so;
\item much of this literature relates concretely or by allusion to
  brains and human cognition, which is the 2370 year old root of
  today's intuitions about ``computing'';
\item all of the visited formalisms and models lend themselves to
  solve practically useful information processing tasks;
\item all satisfy the four necessary conditions that I posited
  (temporality, input/output, cognitive and semantic
  interpretability).
\end{itemize}


Theory development for DC is mature and unifying meta-views are
canonical, which allowed me to inspect this domain almost in its
entirety. In the P and D domains, the untamed diversity of formal
models forced me to limit the coverage to quite narrow subdomains,
namely sampling-based (SPPD) and ODE based modeling methods, respectively. 
Here is a summary of what I saw on my journey. 

\begin{description}
\item[Interrelationships between formalisms.] DC how-formalism (in
  particular, programming languages and abstract models of computing
  machines) can be transparently sorted into the Chomsky hierarchy,
  with ``higher'' formalisms being able to simulate all ``lower''
  ones. Within the highest-level class of formalisms (the
  Turing-equivalent ones), how-formalisms can also be informally 
  ordered according by to how far they abstract away from the 0-1
  switching of binary circuits toward cognitive interpretability.
  What-formalisms (formal logics) can likewise be ordered according to
  their formal intrinsic semantic expressiveness. These ordering
  systems tie ``the theory'' of DC into a unit that fits into a single
  textbook. This crystalline transparency sets intimidating standards
  for attempts to build theories for other sorts of ``computing''.

  The how-models of SPPD are sampling algorithms in cases where
  the SPPD models are physically realized on digital machines; or they
  are mathematical models of stochastic physical processes when the
  target hardware is, for instance, analog spiking neuromorphic
  microchips or DNA computers. The formalisms in which one specifies
  sampling algorithms or neural or DNA random dynamics stem from
  different background mathematics, which makes it hard to analyse
  their mutual relationships; and if one considers sampling
  independently from the mechanics of the generating how-models in the
  abstract light of the resulting generation sequence of sample
  points, there is yet no generally adopted ordering criterion to
  compare sampling sequences (statistical efficiency might serve as
  such a criterion). I do not think however that finding useful
  ordering principles is inherently impossible, an invitation for
  future research. The what-formalism (singular) for SPPD is the
  textbook formalism of probability theory, with the cognitively
  interpretable concept of a probability distribution in its core.
  
  In ODE modeling, the how-formalism is the textbook formalism of
  ODEs. The how-models are concretely spelled-out systems of ODEs
  which specify physical ``computing'' (analog) hardware. Physicists
  would state that any kind of deterministic physical system can be
  described by ODEs to arbitrary degrees of approximation. Thus the
  class of ODE computing models is as diverse as one can think of
  physical computing systems. If the mission statement of MC makes
  sense, this is an open-ended diversity, which makes it seem that
  finding a unifying ordering principle for ODE models is
  impossible. What-models arise in ODE modeling through the
  identification of qualitative phenomena like fixed points,
  attractors, bifurcations and processing modes, to name the ones
  which today are being most widely used as anchors for cognitive
  interpretations. ODEs are a well for an unbounded number other
  qualitiative phenomena that await discovery. This may imply that a
  general what-theory is impossible for ODE modeling.

  It becomes clear that in the DC, SPPD and ODE domains alike we face
  voluminous assortments of formalisms and models. Finding criteria
  and mathematical (meta-) methods to relate them to each other is a
  sign of a domain's integrity and maturity. Only the millenia-old
  field of symbolic computing can claim maturity in this regard.  Here
  is the upshot of this high-altitude flight over the theory
  landscapes in DC, SPPD, and ODE:
  
  \begin{itemize}
    \item DC offers a wealth both of how- and what-formalism which
      are however transparently and comprehensively interrelated.
    \item SPPD has a single ``master'' what-formalism and a yet
      unsorted multiplicity of how-formalisms and models.
    \item Conversely, ODE modeling has a single how-formalism but
      what-formalizing may be principally open-ended and un-unifiable.
  \end{itemize}

\item[Formal semantics.] Formal semantic theories can only be stated
  for the relationship between two formalisms. In our case this means
  that they give mathematical accounts of the ``meaning''
  relationships between cognitively interpretable constructs in
  what-formalisms and formal correlates of objects, facts, processes
  etc.\ in mathematical models of a computing system's
  environment. Such semantic relations create bridges between segments
  \textbf{a} and \textbf{c} in Figure \ref{figGrandSchema}. I find it
  a mandatory component of any theory of ``computing'' that it allows
  one to express how what happens ``cognitively'' inside a computing
  system, relates to conditions in its environment.

  Every logic that serves as a DC what-formalism comes equipped with a
  formal semantic. It precisely defines how the formulas that can be
  written down in the logic (segment \textbf{a}) become
  \emph{interpreted} in formal models (in \textbf{c}) of
  environments. These formal models are cast as certain \emph{richly
    structured sets} called $S$-structures. Formal logics offer no
  mathematical account of the interface boundary \textbf{b} between a
  computing system and its environment. In the view of logicians, the
  symbols and expressions written down in a logic formalism are
  directly \emph{denoting} something in the $S$-structures, without an
  intermediary exchange of ``data'' or ``signals''. When physical
  digital computers (segment $\alpha$) are used in the physical world
  $\gamma$, it is up to the physical (human or automated) user of the
  computer to establish a physical intrinsic semantic linkage between
  the computer and its task environment by providing appropriate input
  and interpreting the computer's output appropriately. Formal logic
  is blind to the interface boundary problem and cannot capture what
  ``appropriately'' means.

  In fundamental contrast, probability theory (the singular
  what-formalism in all probabilistic modeling, not only in SPPD)
  captures ``data'' or ``signals'' in its core constructs,
  namely random variables and sample spaces. Random variables are the
  formal correlate of measurement or observation procedures and
  apparatuses. Sample spaces are sets made from the data values that
  these observation procedures may deliver; they thus perfectly
  coincide with segment \textbf{b} in Figure
  \ref{figGrandSchema}. Like in logics, the formal models of the
  environment are richly structured sets called probability spaces,
  which additionally are endowed with a probabilty measure. In the
  frequentist understanding of probability, this probability measure
  reflects physical randomness. Probability spaces constitute segment
  \textbf{c}. Random variables can be regarded as semantic operators
  in that they connect segments \textbf{a} with \textbf{c} through
  \textbf{b}. However, this connection is unidirectional. While random
  variables capture the \emph{input} given by the environment to
  inform the shaping of probability distributions which are modeled in
  \textbf{a}, there is no provision for capturing any impact which an
  \emph{output} from a probabilistic ``reasoning'' process modeled by
  the how-formalism in \textbf{a} could have on the probability space. 

  In ODE modeling the semantic situation is again fundamentally
  different. In dynamical systems modeling the computing system and
  its environment are seen as coupled subsystems within a single
  dynamical agent-environment system. Both subsystems \textbf{a} and
  \textbf{c} are described through ODEs. They are bidirectionally
  coupled through shared variables which constitute the interface
  boundary \textbf{b}. The mathematical substrate for how-modeling
  both the computing ``agent'' as well as its environment are vector
  fields --- there is no difference in mathematical kind between the
  modeled inner and outer worlds.  In dynamical systems oriented
  research where agent-environment interactions are studied, these
  interactions are not understood as ``semantic''; in the
  epistemological view of constructivism and enactivism these models
  are even advanced as anti-semantic, denying that inside an agent
  there are mental ``representations'' of outward givens. --- However,
  if one adopts the view which I take in this article, namely to
  locate the formal intrinsic semantic relationship between
  what-models and models of the environment, the semantic question
  arises again. Concretely, one would ask how cognitively
  interpretable dynamical phenomena (attractors, bifurcations,
  modes... in segment \textbf{a}) can be formally connected to
  environment models. This is unchartered territory.

  These findings reveal that there are indeed very different ways to think
  about semantics. ``Semantics'' is not a clearly defined concept
  and every philosopher worth his/her salt will understand this term
  in a different way. This opens many degrees of freedom for
  developing MC theories. Here is the upshot of our high-speed drive
  through the semantics challenge in DC, SPPD, and ODE:

  \begin{itemize}
  \item In DC and SPPD, accounts of intrinsic semantics are
    firmly established in fundamental mathematical definitions. They
    specify how cognitively interpretable formal constructs in models
    of ``computing'' system relate to formal models of their
    environments. Nothing comparable is yet available for ODE
    modeling.
  \item In DC and SPPD, the mathematical ``substrate'' of environment
    models is different in kind from the substrate of the cognitively
    interpretable constructs in agent models, namely set-theoretic
    structures versus symbolic configurations and probability
    distributions, respectively. In ODE modeling, both sides are made
    of the same mathematical material, namely vector fields. 
  \item SPPD and ODE modeling comprises canonical constructs for the
    interface boundary (data, signals) between a computing system and
    its environment; logic doesn't.    
  \end{itemize}

\item[Formal time.] Regardless how one understands the nature of
  ``computing'', one thing seems inevitable: physical computing
  systems need time to ``compute''. Any formal theory of any kind of
  computing should also model the computing system's time, certainly
  in its how-formalisms and models. They are the springboards from
  which system engineers jump across the modeling mirror and build
  machines, guided by formal how-models in the back of their minds.  In
  the three domains that I inspected, time is modeled in interestingly
  different ways in how-models.  Even more interesting are the
  differences between DC, SPPD and ODE with regard of how the physical
  time that is modeled in how-models relates to the ``conceptual''
  time that is (or is not) formalized in what-models.

  In DC how-formalisms, time enters in the form of update steps where
  one symbolic configuration is transformed into the next. These steps
  unfold into nonzero-duration increments of physical time on the
  physical digital machines. The physical duration can be longer or
  shorter (depending on the clock speed and the degree of CPU-internal
  parallelization) and it is not modeled in typical how-formalisms
  (real-time operating systems excepted). When one sees these update
  steps in the light of logic-based what-formalisms --- the light
  shining from Aristotle and Turing --- they are not seen as temporal
  at all, but as inferential. The next symbolic configuration follows
  logically from the previous.  The verb ``follow'' is fascinatingly
  ambiguous, with one of its meanings being temporal succession and
  another one being logical implication. There are more words which
  have a temporal and a logical-inferential side, for instance
  ``consequence'', ``conclude''. Seen from this angle, the history of
  DC could be summarized like this: First it took philosophers and
  mathematicians almost 2300 years to strip logical inference off from
  the physical brain's physical time, a process which became finalized
  in Tarski's reconstruction of logical implication in terms of static
  inclusion relations between classes of $S$-structures. Then
  temporal succession was re-introduced by Turing in the form of an
  ordered sequence of discrete symbolic configuration update
  ``steps''. Ultimately, electrical engineers and chip manufacturers
  rejoin physical time by realizing these steps within the
  clock cycles of digital microchips.

  In SPPD how-models time can be cast as a discrete sequence of update
  steps (in sampling algorithms destined for execution on digital
  machines), or it can be cast as the continuous time-line
  $t \in \mathbb{R}$ when the sampling process is modeled for use in
  non-digital hardware, in particular in analog spiking neuromorphic
  microchips. In both variants there appears a fundamental difference
  to DC thinking. The cognitively interpretable constructs (the
  probability distributions handled in what-models) need nonzero
  timespans to be grown. The longer a
  sampling process carries on, the more precisely it defines its
  distribution. The cognitively interpretable constructs are smeared
  out over time. If one sees a ``computing'' process as reflecting
  some aspect of a succession of ``mental states'' (which Turing did
  --- and we all stand on his shoulders), then these mental ``states''
  become defined only across time, with their constituting components
  (the representations of distributions) growing, overlapping in time,
  decaying. This is in stark contrast to the DC view where at each
  time point the ``mental state'' is perfectly and completely defined
  by a symbolic configuration, and future configurations do not
  gradually grow out of previous ones but are created in their
  completion immediately and discontinuously.

  ODE how-formalisms (ODE specifications of biological or engineered
  computing systems) use the real line $\mathbb{R}$ as their model of
  time. The temporal evolution $\dot{\mathbf{x}}$ of the state vector
  $\mathbf{x}$ progresses smoothly and with perfect real-valued
  precision. This makes it \emph{possible} for engineers and
  neuroscientists to directly check the adequacy of their ODE
  how-models in matching their physical target systems by measuring
  physical system variables with appropriate measurement
  apparatuses. It also makes it \emph{necessary} for engineers to
  design their analog hardware such that its physical timescales
  correspond to the ones of the model. In the light of own experience
  with analog neuromorphic hardeware, I consider it a decisive
  challenge for future MC engineering to master timescale spreads and
  timescale interactions both in theory and in physical devices and
  systems.  The problem of setting up (hierarchies of) appropriate
  \emph{time constants} is intimately connected to the theme of
  hierarchical organisation of ``computing'' and will be discussed
  below. --- Most of the cognitively interpretable constructs treated
  in what-formalisms and models (attractors, bifurcations, etc.) are
  defined by topological-geometrical phenomena in \emph{phase
    portraits}. ``Speed'' becomes factored out in these analyses:
  phase portraits are made from trajectory lines, and the information
  how ``fast'' the state evolution progresses along these lines is
  discarded. The exception is the phenomenon of modes, which are
  temporally defined --- more about them later. This is a parallel
  with DC and probabilistic modeling, whose cognitively interpretable
  what-constructs (symbolic configurations, distributions) are
  likewise atemporal.

  Human cognition proceeds in time, physical computing systems run in
  time, and we all share a primal intuitive understanding of
  ``time''. One thus would expect a universal, intuitively immediately
  graspable capture of time in ``computing'' theories and
  formalisms. But we find diversity and detachment from intuition.  Here
  is the upshot of our meandering sailing trip through the time
  modeling challenge in DC, SPPD, and ODE:

  \begin{itemize}
  \item Time in how-formalism is cast as a sequence of discrete update
    steps (in DC and in sampling algorithms) or as continuous (in dynamical
    systems models of neural sampling processes and ODEs).
  \item Matching formal time in how-formalism with physical time in
    the modeled computing systems is arbitrary in DC, and well-defined
    and measurable in ODE. In SPPD both occurs depending on whether
    the sampling becomes physically instantiated on digital or
    non-digital physical systems.
 \item The cognitively interpretable constructs
    that are commonly expressed in the what-models of DC, SPPD and ODE
    are allmost all a-temporal, which is somewhat amazing since human
    cognition is temporal. The one exception are modes in ODE models,
    which are inherently temporal phenomena. They will play an
    important role  in my suggestions for starting a theory of MC in
    the concluding section of this article. 
  \end{itemize}

\item[Hierarchical structuring of formal constructs.] A characteristic
  of human cognitive processing is \emph{compositionality}: we can
  compound syllables into words into phrases, bind noses, eyes and
  mouths into faces, plan complex plans that unfold in cascades of
  sub-plans to reach sub-goals --- we can think \emph{complex}
  thoughts. This ability has been claimed constitutional for human
  intelligence, and it seems natural to request the same from any
  full-scale ``computing'' system and its theoretical
  models. 

  In DC how-formalisms, hierarchies appear in two main ways. First,
  the symbolic configurations, which are stepwise constructed when
  how-models are ``executed'', typically are organized as
  hierarchically nested composites. Second, this \emph{syntactic}
  compositionality of symbolic configurations is typically tied in
  with a \emph{procedural} hierarchical organization of ``runs'' of
  programs or formal machine models: symbolic substructures are built
  within program ``loops'' or by calling ``subroutines'', with the
  effect that substructures correspond to sub-intervals in processing
  time. Writing a nontrivial computer program amounts  to breaking down
  the global input-to-output functionality imposed by the given task
  into a nested sequence of intermediate goals and subgoals. Much sweat
  is spent by students in software engineering classes to
  acquire this skill. --- The symbolic expressions which are written
  down in DC how-formalisms (that is, logics) typically contain nested
  functional expressions which are encodings of (parts of) the
  symbolic configurations in how-formalisms.

  In SPPD what-formalisms (and probabilistic models of ``computing''
  in general), the primary cognitively interpretable constructs are
  distributions. Distributions can be sees as compositional in several
  ways. First, a number of popular stochastic spiking neural network
  architectures are layered, in analogy to the peripheral-to-central
  processing organization in human brains. When used in (for example)
  in face recognition tasks, samples collected from low-level neurons
  are considered to represent local visual features (like colors,
  edges or dots) of input images, whose information becomes
  increasingly combined and globalized in higher layers (from edges to
  contour segments to eyes to faces). The composition operation here
  is different from DC. In intuitive terms, compound symbolic
  configurations in DC are put together like Lego bricks. Higher-level
  distributions in spiking neural architectures are statistically
  determined from the sampling dynamics in the lower layer. This could
  be likened to an argumentation process where a stream of lower-level
  ``arguments'' integrates up into higher-level ``beliefs''. Second,
  in \emph{Bayesian} models of cognitive information processing,
  ``higher'' distributions arise as \emph{hyperdistributions}, that
  is, distributions of distributions. Third, a fundamental textbook
  operation on distributions is to combine them into \emph{products}
  (where the component distributions remain statistically independent)
  or \emph{joint distributions} (wherein the ``component''
  distributions interact and become statistically dependent on each
  other). Conversely, high-dimensional distributions can sometimes be
  more or less precisely \emph{factorized} into low-dimensional
  component distributions.

  In ODE systems, a (in my view \emph{the}) key to hierarchical
  structuring of qualitative phenomena are timescale hierarchies. In
  multiple-timescale ODE systems, the dynamics evolves through a
  sequence of hierarchically nested \emph{modes} in alignment with the
  nested characteristic timescales of mode-controlling system
  variables. According to a widespread view in the cognitive and
  neurosciences, the hierarchy of timescales is a mirror of the
  compositional hierarchy of cognitive ``representations''. This leads
  to the pervasive idea that cognitive architectures are layered
  structures in which ``higher'' processing layers evolve more slowly
  than ``lower'' layers. When there are both bottom-up and top-down
  couplings between neighboring layers, mathematical analysis becomes
  challenging. Furthermore, in biological brains and advanced
  technical control systems no linear ordering of processing layers
  exists. Subsystems interact not only along a single top-down /
  bottom-up direction, but are also coupled laterally or
  diagonally. Relative timescales may change, slow subsystems turning
  into fast ones and vice versa. Such phenomena are hardly understood.

  The essence of compositional hierarchies in L, P, D is that elements
  that are higher in the hierarchy are made from elements from
  lower layers. This ``made from'' relation, however, can mean quite
  different things in different approaches to modeling
  ``computing''.  Here is the upshot of our short dash into the
  thickets of hierarchical structuring phenomena of DC, SPPD, and ODE:

  \begin{itemize}
  \item The symbolic configurations in DC how- and what-formalisms are
    structured in a static-syntactic Lego-brick kind of hierarchical
    compositionality.
  \item Probability distributions can be seen as hierarchically
    structured in several ways, all of which are not ``syntactically''
    defined but can be more appropriately understood by observing that
    component distributions shed some of their statistical information
    into the compound distribution.
  \item A key to hierarchic organization in complex dynamical systems is a
    hierarchy of timescales, which induce a hierarchically nested
    sequence of processing modes. 
  \end{itemize}

  
\end{description}

The title of this section is \emph{Staking out the ``computing''
  theory landscape}. I could cover the dominion of digital computing
almost in its entire extension, though of course with simplifications and
omissions. This was possible because theoretical (digital) computer
science is mature, unified and canonized; thus all I had to
do is to map \emph{the} theory to the organigram of
Figure \ref{figGrandSchema}. For probabilistic and dynamical systems
oriented models of ``computing'', unified meta-views are not in
sight. In order to not get lost I selected small sectors of them,
namely sampling-based computational methods to represent probability
distributions, and ODE modeling. But even within this limited angle of
vision, landmarks and signposts came into view which invite us to
explore ``computing'' in many more directions than those of the
digital-symbolic paradigm. Here is my personal grand total of this first
expedition into the landscape of ``computing'':

\begin{itemize}
\item The way of how one \emph{can} conceptualize ``computing'' is
  decisively pre-shaped by the choice of mathematical ``substrate''
  formalism (here: logic, probability theory, dynamical systems).
\item Different conceptualizations of ``computing'' grow
  around different aspects of human cognition (logical inference,
  probabilistic reasoning and degrees of belief, continuous
  sensor-motor coordination).
\item Digital computing will forever remain the emperor over the
  entire ``computing'' realm in the sense that digital computer
  programs can carry out logic inference; logic (together with set
  theory) can express all of mathematics; all formal how-models of
  ``computing'' are mathematical;  hence digital computers can simulate
  all other formal procedural specifications of
  ``computing''. However, this emulation can become prohibitively
  inefficient with regards to runtimes, energy consumption and
  microchip complexity. 
\item Procedural formalisms (I called them how-formalisms) are the
  springboards for engineers from which they jump across the modeling
  mirror, building physical computing systems which realize the formal
  specifications.  Depending on the chosen mathematical substrate,
  different limitations and opportunities for physical designs
  arise. Digital system engineers \emph{must} build hardware based on
  finite-state switching operations and memory mechanisms to stably
  store switching states for very long times. Once they know how to
  build such machines, they \emph{can} capitalize on the full powers of
  symbolic computing theory. System engineers informed by sampling
  how-formalisms \emph{must} find ways to harness physical
  stochasticity. Once they master this task, they \emph{can} build
  machines which realize already existing, general models of
  probabilistic inference (graphical models, in particular Boltzmann
  machines). At present, physical randomness has been made exploitable
  for sampling only in limited ways in DNA computing (note that
  quantum computing exploits randomness not by sampling and was not
  covered in this article). System engineers guided by ODE models of
  cognitive processing \emph{should} learn to realize an ever growing
  repertoire of ODE models in physical dynamics (asymptotic goal: find
  ways to implement \emph{any} ODE specification). Then they
  \emph{could} build machines which re-play the cognitive mechanisms
  that have been discovered and will be discovered, in the wider
  cognitive and neurosciences, as qualitative
  phenomena in dynamical systems. They could achieve even more if a
  dynamical systems understanding of ``computing'' disengages itself
  from its current reliance on ``brain-inspired'' neuromorphics.
\item How-theories of ``computing'' include, implicitly or explicitly,
  a formal model of \emph{time}, which in turn co-determines which
  cognitive operations can be captured in corresponding
  what-formalisms, and how. The importance of analysing how
  ``computing'' processes are structured in time is, in my opinion,
  largely under-appreciated. The same could likely be said about
  \emph{space}, a theme which I decided to leave out in this
  article. Temporal and spatial phenomena are closely coupled in
  physical systems, a given that demands an extensive discussion
  which I postpone to another occasion.  
\end{itemize}

\section{Fitting modal computing into the
  landscape}\label{secConcrete}

 Despite the limited scope
of reconnaissance in the previous section I hope that what I spotted
gives some helpful orientation in the search for a theory of modal
computing (MC). In this concluding section I want to propose some ideas
toward this end. 

The mission for MC is to transform the not-so-new but still vague idea
of exploiting ``physics directly'' into a solid engineering
discipline. I advanced an even stronger version of this idea --- the
mission is to harness \emph{any} kind of physical phenomenon which
supports ``computing''.  This will require a collaboration between
contributors who today rarely work together --- material scientists,
theoretical physicists, microchip engineers, neuroscientists,
cognitive scientists, AI and machine learning experts, computer
scientists, mathematicians and epistemologists. They can only start
talking with each other after they have agreed on a shared language,
and they can only precisely understand what they think they are
talking about after having developed a mathematical foundation
underneath this language.

In my opinion, which I tried to substantiate in this serpentine
article, any theory about any kind of physical systems can qualify as
a theory of ``computing'' only if core constructs in that theory can
be linked to human cognitive processing. This view is authorized by
2370 years of intellectual labour, starting with Aristotle's
syllogistic logic and not ending with Alan Turing who explicitly
equated the symbolic configurations in Turing machines with ``states
of mind''.

In my review of the D, P and L domains I placed the formal
correspondents of cognitive entities into the center of what I called
what-formalisms (in segment \textbf{a} in Figure
\ref{figGrandSchema}), and inspected the formal semantic links which
connect those formal constructs with formal models of environments
(segment \textbf{c}). Across the modeling mirror, what is formally
represented in models of computing systems and models of environments
should have reflections in the corresponding physical computing system
(in $\alpha$) and its physical environment (in $\gamma$). Furthermore,
the physical computing system should enjoy physical semantic relations
with its environment which make the diagram
\begin{center}
\includegraphics[width=0.2\textwidth]{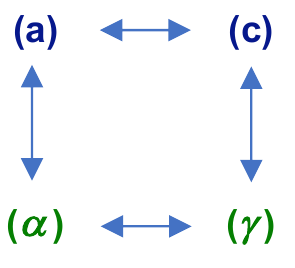}
\end{center}
\noindent commute. This implies that there should be a correspondence
between the cognitively interpretable constructs in what-formalisms (proceduralized in how-formalisms) on the one hand, and
phenomena in physical computing systems on the other hand. This is in
fact what we could observe in digital computing (discrete symbol
operations corresponding with binary switching dynamics), in
sampling-based probabilistic models (formal sample points
corresponding to, for example, spikes in neural substrates or DNA
sniplets in DNA computing), and in ODE modeling (for instance,
periodic attractors realized in oscillatory electronic circuits).

Thus, when we want to capitalize on \emph{any} kind of
computing-enabling physical phenomenon in MC machines, we
have to address the question, \emph{What is it in human cognition that
  allows us to mentally represent ``any'' kind of physical
  phenomenon?} Well, I am afraid that human intelligence cannot grasp
``any'' kind of physical phenomenon. This version of an MC mission
statement seems too strong. By reversing the direction of the argument
we obtain a more moderate question: \emph{What physical phenomena can
  be cognitively grasped?} But this framing still is too wide. In the
light of findings in the previous section, we should more narrowly ask
this question: \emph{What temporal physical phenomena, which can be
  coupled into increasingly compounded complex phenomena, can be
  cognitively grasped?} or its mirror twin: \emph{What cognitive
  phenomena make us perceive and think about physical processes as
  being constituted of coupled sub-processes?}

I want to illustrate this abstract question with one concrete
example. Attach LEDs to some joints and extremity ends of a human
volunteer. Let this person move in an unlit room such that only the
light traces of these LEDs are visible. Then a human observer
can identify whether the volunteer walks, runs, jumps, waltzes, or
engages in any of hundreds human motion patterns. The observer can
even tell who the performing volunteer is, even from just
monitoring a single short walk, provided it is a personal
acquaintance. The observer can also decompose the overall dynamic
pattern presented by all the LEDs into subpatterns, for instance
focussing on the right arm's motion. --- This is an example from
visual sensory processing. But the scope of our twin question is far
wider and could be filled with examples from other sensory modalities;
not only from perception but also from action generation; or from
the entire mental experience of being present in a dynamic
environment; or even from a mathematician sitting still in deep
thought in front of a white sheet of paper --- mathematical thinking
arguably being just another way of re-experiencing dynamical
situatedness \citep{LakoffNunez00}.   

If we had a formal theory which allowed us to state and work out this
twin question in appropriate and precise abstraction, all that would
remain is to team up with engineers and material scientists and start
building machines which instantiate the formal constructs of our
theory. This would be similar to, but more general than, informing
electronics engineers to build analog computers which can instantiate
a number of elementary mathematical operations like multiplication or
integration.

Devising of such a theory does not start from a blank slate. Cognitive
scientists have since long been investigating complex, gradually
morphable mental representations of dynamical phenomena, developing
comprehensive theory frameworks like \emph{fluid
  concepts} \citep{Hofstadter95} or \emph{radial
  categories} \citep{Lakoff87}, or exploring complex motion pattern
representations \citep{Blasingetal09, Tervoetal16}. These are random
pointers; a structured survey remains to be done.

Theories in cognitive science, in particular in its experimental
branches, are however often articulated only in natural English and
corroborated by computer simulations. If one searches for inspiration
from mathematically worked-out theories, one can find it in places
outside the core cognitive sciences. I personally have felt instructed
by mathematical models in (human and animal) motion science where one
objective is to formally describe complex motor patterns and analyze
how they can be controlled \citep{HoganFlash87, ThoroughmanShadmehr00,
  dAvelleetal03}. My LED-tracing example was borrowed from recent work
in this line \citep{Landetal13}. Grenander's \emph{pattern theory},
especially in the transparent rendering of David Mumford
\citep{Mumford94, Mumford02}, offers a rich and thoroughly formal
account of how (primarily spatial / visual) ``patterns'' which are
emerging in complex physical systems can be generated, compounded,
transformed and encoded. A classical subfield of AI, \emph{qualitative
  physics} \citep{Forbus88} (closely related: \emph{naive physics},
\emph{qualitative reasoning}) explores logic-based formalisms which
capture the everyday reasoning of humans about their mesoscale
physical environment. Insights gained in the fields of \emph{emergent
  computation} \citep{Forrest1990a} steer attention to the powers of
collective phenomena in dissipative systems, where macrolevel
phenomena ``self-organize'' from the interactions of microlevel
components. Machine learning and data mining methods for detecting
\emph{concept drift} \citep{Gamaetal13} offer statistical
characterizations of how data streams change qualitatively over time,
including recent methods which exploit hierarchical structuring of
distributions \citep{Hammoodietal18}. Finally, recent propositions to
develop a theory of \emph{stream automata} \citep{Endrullisetal20} aim
at extending the classical theory of finite-state automata to infinite
data stream processing.  These and other sources of mathematical
inspiration remain to be surveyed and connected.

At this moment I do not see a single, uniquely compelling
\emph{ansatz} which holds promise to become worked out into a formal
theory that could answer the twin question posed above. In a separate
manuscript (in preparation) I describe in more detail why I think that
such a system of interconnected formal theories can be rooted in the
concept of dynamical modes, which generalize the bi-stability
modes of digital switching transistors.

I believe that we are facing a quite fundamental challenge, and that
at the core we are even lacking an adequate mathematical
language. Newton and Leibniz devised calculus to capture
\emph{continuous motion}. Kolmogorov and his predecessors developed
probability theory to capture the \emph{information conveyed by
  empirical observations}. Tarski and his predecessors cast logics in
its final shape in order to capture \emph{rationally derivable
  truth}. If I were pressed to condense that twin question of MC into
a similarly momentous three-word phrase I would say that we have to
capture \emph{gradual qualitative change} (which I want to formally
cast in a concept of dynamical modes). I believe that this asks
from us to discover a profoundly new mathematical language, a new
branch in the tree of mathematics which grows between probability (for
``gradual''), logic (for ``qualitative'') and dynamical systems (for
``change''). I sometimes tell my students that I hope to live to the
day when one among them finds it.

\vspace{0.5cm}

\textbf{Acknowledgements.}
The authors would like to acknowledge the financial support of the CogniGron research center and the Ubbo Emmius Funds (Univ. of Groningen).




\end{document}